\documentstyle[multicol,amssymb,pra,aps,eqsecnum,epsf]{revtex}

\newtheorem{proposition}{Proposition}
\newcommand{\Proof}{{\smallskip\bf \noindent Proof. }}
\newcommand{\EndProof}{{\hfill{$\square$} \vskip4pt}}

\begin{document}
\draft

\title{Quantum stochastic models of two-level atoms and electromagnetic cross
sections. }

\author{Alberto\ Barchielli}
\address{Dipartimento di Matematica, Politecnico di Milano,
 Piazza Leonardo da Vinci 32, I-20133 Milano, Italy
\\ and  Istituto Nazionale di Fisica Nucleare, Sezione di Milano.
 E-mail: barchielli@mate.polimi.it}

\author{Giancarlo\ Lupieri}
\address{Dipartimento di Fisica, Universit\`a degli Studi di Milano,
 Via Celoria 16, I-20133 Milano, Italy
\\ and  Istituto Nazionale di Fisica Nucleare, Sezione di Milano}

\date{\today}
\maketitle

\begin{abstract}
Quantum stochastic differential equations have been used to describe the
dynamics of an atom interacting with the electromagnetic field via
absorption/emission processes. Here, by using the full quantum stochastic
Schr\"odinger equation proposed by Hudson and Parthasarathy fifteen years ago,
we show that such models can be generalized to include other processes into the
interaction. In the case of a two-level atom we construct a model in which the
interaction with the field is due either to absorption/emission processes
either to direct scattering processes, which simulate the interaction due to
virtual transitions to the levels which have been eliminated from the
description.

To see the effects of the new terms, we study various types of cross sections
for the scattering of monochromatic coherent light. We obtain formulas giving
the total, the elastic and the inelastic cross sections as functions of the
frequency and the intensity of the stimulating laser and the fluorescence
spectrum as a function also of the frequency of the scattered light. The total
cross section, as a function of the frequency of the stimulating laser, can
present not only a Lorentzian shape, but the full variety of Fano profiles;
intensity dependent widths and shifts are obtained. The fluorescence spectrum
can present complicated shapes, according to the values of the various
parameters; when the direct scattering is not important the usual symmetric
triplet structure of the Mollow spectrum appears (for high intensity of the
stimulating laser), while a strong contribution of the direct scattering
process can distort such a triplet structure or can even make it disappear.
\end{abstract}

\begin{multicols}{2}
\narrowtext

\section{Introduction}\label{intro}
Quantum stochastic calculus (QSC) \cite{1HP84,GarC85,Gar,Partha92}, a
noncommutative analog of the classical Ito's stochastic calculus, revealed to
be a powerful tool to construct mathematical models of quantum optical systems
\cite{Gar,Gar86,Bar87,KW88,AMW88,LRW88,MRW88,CW88,WM94} and to develop a theory
of photon detection \cite{BarL85,Bar86,Bar90,BPag}. Just at the beginning of
QSC, Hudson and Parthasarathy proposed a quantum stochastic Schr{\"o}dinger
equation for quantum open systems \cite{1HP84,Partha92}. Such an equation has
found applications in quantum optics, but not in its full generality
\cite{Bar87,Bar90,Gar}. It has been used to give, at least approximately, the
dynamics of photoemissive sources such as an atom absorbing and emitting light,
or matter in an optical cavity, which exchanges light with the surrounding free
space. But in these cases the possibility of introducing the so called gauge
(or number) process in the dynamical equation has not been considered; roughly
speaking, the gauge process is a quadratic expression in the field operators
which preserves the number of quanta, but changes their wave functions. In this
paper we want to show, in the case of the simplest photoemissive source, namely
a two-level atom stimulated by a laser, how the full Hudson-Parthasarathy
equation allows to describe in a consistent way the scattering of the light by
the atom not only through the absorption/emission channel, but also through
another process which can be called ``direct scattering'', which can be
included in the interaction via a term containing the gauge process. When the
atom is approximated by a two-level system, the introduction of an interaction
term which preserves the number of photons allows to simulate also the
scattering processes involving virtual transitions to states different from the
two ones responsible of the real absorption/emission process.

So, a first aim is to show how the full Hudson-Parthasarathy equation is able
to give a reasonable and rich model for the dynamics of an atom interacting
with the electromagnetic field. A second one will be the study of the elastic,
inelastic and total cross sections for the scattering of monochromatic coherent
light by the atom. The resulting line-shapes are very interesting. For
instance, the dependence of the total cross section on the frequency of the
stimulating laser can present not only a Lorentzian shape, but the full variety
of Fano profiles $(\!\!{}$\nobreak\cite{Fano}, \cite{CT92} pp.\ 61--63).
Moreover, the dependence of the line shape on the intensity of the stimulating
laser is computed and power broadening and intensity dependent shifts are
found. The study of the inelastic cross section, instead, shows possible
modifications to the known triplet structure of the fluorescence spectrum
\cite{Moll}. Some preliminary results on the total cross sections where
reported in \cite{BarL98}.

\subsubsection*{Fock space and QSC}
Let us recall some notions of QSC and the Hudson-Parthasarathy equation; this
is just to fix our notations, while for the proper mathematical definitions and
the rules of QSC we refer to the book by Parthasarathy \cite{Partha92}. We
denote by ${\cal F} = {\cal F}({\cal X})$ the Boson Fock space over the
``one-particle space'' ${\cal X} = {\cal Z} \otimes L^2({\mathbb R}_+) \simeq
L^2({\mathbb R}_+; {\cal Z})$, where ${\cal Z}$ is another separable complex
Hilbert space. A vector $f$ in $\cal X$ is a function from ${\mathbb R}_+$ into
$\cal Z$; we fix a c.o.n.s.\ $\{e_i,\ i\geq 1\}$ in ${\cal Z}$ and we denote by
\begin{equation}\label{1.3}
 f_j(t) = \langle e_j| f(t) \rangle
\end{equation}
the components of a vector $f(t)$ in $\mathcal Z$. The Fock space $\mathcal F$
is spanned by the exponential vectors $E(f)$, whose components in the
$0,1,\ldots,k,\ldots$ particle spaces are
\begin{equation}\label{1.5}
E(f) = \left(1,f,(2!)^{-1/2}f\otimes f,\ldots,(k!)^{-1/2} f^{ \otimes k},
\ldots \right),
\end{equation}
$f \in \mathcal X$; the inner product between two exponential vectors is given
by
\begin{eqnarray}\nonumber
 \langle E(g) | E(f) \rangle &=& \exp\, \langle g | f \rangle
\equiv \exp \biggl[ \int_{-\infty}^{+\infty} \langle g(t)|f(t) \rangle
\,{\mathrm d} t \biggr]
\\ \label{1.6}
&\equiv& \exp \biggl[ \sum_j \int_{-\infty}^{+\infty} \overline
{g_j(t)}\,f_j(t)  \,{\mathrm d} t \biggr],
\end{eqnarray}
where an overline means complex conjugation, and we get normalized vectors by
defining
\begin{equation}\label{1.7}
e(f)= \exp\left(-\textstyle\frac{1}{2}\|f\|^2\right) E(f)\,.
\end{equation}

The annihilation, creation and gauge (or number) processes are defined by
\begin{eqnarray}\nonumber
A_j(t) E(f) &=& \int_0^t f_j(s)\,{\mathrm d} s\, E(f)\,,
\\ \label{1.8}
\langle E(g)| A^\dagger_j(t)E(f)\rangle &=& \int_0^t \overline{g_j(s)}\,
{\mathrm d} s \, \langle E(g)|E(f)\rangle\,,
\\ \nonumber
\langle E(g)|\Lambda_{ij}(t)E(f)\rangle &=& \int_0^t \overline{g_i(s)}f_j(s)\,
{\mathrm d} s \, \langle E(g)|E(f)\rangle\,.
\end{eqnarray}
Eqs.~(\ref{1.8}) allow to write formally
\begin{eqnarray}\nonumber
A_j(t) &&{}= \int_0^t a_j(s) \,{\mathrm d} s\,,
\\ \label{1.4}
A_j^\dagger(t) &&{}= \int_0^t a_j^\dagger(s) \,{\mathrm d} s\,,
\\ \nonumber
\Lambda_{ij}(t) &&{}= \int_0^t a_i^\dagger(s)a_j(s) \,{\mathrm d} s\,,
\end{eqnarray}
where $a_j(t)$, $a_j^\dagger(t)$ are usual Bose fields, satisfying the
canonical commutation rules
\begin{equation}\label{1.9}
 [a_j(t),a_i(s)] = 0\, , \qquad
[a_j(t),a_i^{\dagger} (s)] = \delta_{ji}\, \delta(t - s)\,,
\end{equation}
and whose coherent vectors are the normalized exponential vectors:
\begin{equation}\label{1.10}
 a_j(t)\,e(f) = f_j(t)\, e(f)\,.
\end{equation}
In particular the vector $e(0)\equiv E(0)$ is the Fock vacuum.

The Bose fields introduced here represent a good approximation of the
electromagnetic field in the so called \emph{quasi-monochromatic paraxial
approximation} \cite{Yuen,Bar90}. Now, ${\cal F}$ is interpreted as the Hilbert
space of the electromagnetic field; $A^\dagger_j(t)$ creates a photon with
state $e_j$ in the time interval $[0,t]$, $A_j(t)$ annihilates it,
$\Lambda_{jj}(t)$ is the self\-ad\-joint operator representing the number of
photons with state $e_j$ in the time interval $[0,t]$ and
\begin{equation}\label{1.1}
N(t)= \sum_j \Lambda_{jj}(t)
\end{equation}
is the observable ``total number of photons entering the system up to time
$t$''. Moreover, in the approximation we are considering, the fields behave as
monodimensional waves, so that a change of position is equivalent to a change
of time and viceversa. If we forget polarization, the one-particle space ${\cal
Z}$ has to contain only the degrees of freedom linked to the direction of
propagation \cite{Bar-direct}, so that we can take
\begin{eqnarray}\nonumber
{\cal Z} &=&  L^2 \big(\Upsilon,\,\sin \theta\, {\mathrm d} \theta\,{\mathrm d}
\phi \big)\,,
\\
\Upsilon &=& \{0\leq
  \theta \leq \pi, \ 0
  \leq \phi < 2\pi\} \,;\label{3.22}
\end{eqnarray}
the angular coordinates $(\theta,\phi)$ represent the direction of propagation.
Now, a vector $f$ in the one-particle space $\cal X$ can be identified with a
function $f(\theta, \phi, t)$ such that $\int_0^{+\infty} {\mathrm d} t
\int_0^\pi {\mathrm d} \theta \, \sin \theta \int_0^{2\pi} {\mathrm d} \phi
\left| f(\theta, \phi, t)\right|^2 < +\infty$.

In QSC integrals of ``Ito type'' with respect to ${\mathrm d} A_j(t)$,
${\mathrm d} A_j^\dagger(t)$, ${\mathrm d} \Lambda_{ij}(t)$ are defined. The
main practical rules to manipulate ``Ito differentials'' are the facts that
${\mathrm d} A_j(t)$, ${\mathrm d} A_j^\dagger(t)$, ${\mathrm d}
\Lambda_{ij}(t)$ commute with anything contain the fields only up to time $t$
and that the products of the fundamental differentials satisfy
\begin{eqnarray}\nonumber
&&{\mathrm d} A_j(t)\, {\mathrm d} A_i^{\dagger}(t) = \delta_{ji}\, {\mathrm d}
t\,,
\\ \nonumber
 &&{\mathrm d} A_j(t)\, {\mathrm d} \Lambda_{ki}(t) = \delta_{jk}\, {\mathrm d}
 A_i(t)\,,
 \\ \label{1.11}
&&{\mathrm d}\Lambda_{ji}(t)\, {\mathrm d} A_k^\dagger(t) = \delta_{ik}\,
 {\mathrm d} A_j^\dagger(t)\,,
 \\ \nonumber
&&{\mathrm d} \Lambda_{ji}(t)\, {\mathrm d} \Lambda_{lk}(t) = \delta_{il}\,
{\mathrm d} \Lambda_{jk}(t)\,,
\\ \nonumber
 &&{\mathrm d} A_i^\dagger(t)\, {\mathrm d} A_j(t) =
{\mathrm d}\Lambda_{ki}(t)\, {\mathrm d} A_j(t) = {\mathrm d} A_k^\dagger(t)\,
{\mathrm d} \Lambda_{ji}(t) = 0\,;
\end{eqnarray}
all the products of ${\mathrm d} A_j(t)$, ${\mathrm d} A_j^\dagger(t)$ or
${\mathrm d} \Lambda_{ij}(t)$ with ${\mathrm d} t$ vanish.

\subsubsection*{The evolution equation}
Let ${\cal H}$ be a separable complex Hilbert space (the system space) and let
$R_i$, $i\geq 1$, $S_{ij}$, $i,j\geq 1$, $H$ be bounded operators in ${\cal H}$
such that $H^\dagger =H$, $\sum_i R_i^{\dagger } R_i$ is strongly convergent to
a bounded operator, and $\sum_{i,j} S_{ij} \otimes |e_i \rangle \langle e_j|=
S\in {\cal U}({\cal H}\otimes {\cal Z})$ (unitary operators in ${\cal H}
\otimes {\cal Z}$); we set also
\begin{equation}\label{1.12}
K=H - \frac{{\mathrm i}}{2} \sum_j R_j^{\dagger }R_j\,.
\end{equation}
Then $(\!\!{}$\cite{Partha92} Theor.\ 27.8 p.\ 228) there exists a unique
unitary operator-valued adapted process $U(t)$ satisfying $U(0) =\openone$ and
\begin{eqnarray}
{\mathrm d} U(t) = {}&&\bigg\{ \sum_j R_j \,{\mathrm d} A_j^\dagger(t) +
\sum_{i,j} \left(S_{ij}- \delta_{ij}\right) {\mathrm d} \Lambda_{ij}(t)
\nonumber
\\ \label{1.2} &&{}- \sum_{i,j} R_i^{\dagger } S_{ij }\,{\mathrm d}
A_j(t) -{\mathrm i} K\,{\mathrm d} t \bigg\} \, U(t)\,.
\end{eqnarray}

The operator $U_t$ will be the evolution operator for the atom-field system, in
the interaction picture with respect to the free dynamics of the field. In
order to describe a two-level atom, we take ${\cal H}= {\mathbb C}^2$; then, to
fix the model, we have to determine the atomic operators $H$, $R_i$, $S_{ij}$
on the basis of physical considerations. In the next section we shall require:
(a) the existence of a ground state to which the atom decays by emitting at
most one photon when it is not stimulated, (b) a balance equation [Eq.\
(\ref{2.15})] between the numbers of ingoing and outgoing photons when there is
some coherent source. This suffices to determine the structure of the atomic
operators [Eqs.\ (\ref{2.11}), (\ref{2.16})].

\subsubsection*{Contents}
As said before, in Section \ref{model} we fix the model by physical
considerations; here, a central role is played by a balance equation saying
that the mean number of outgoing photons plus the mean number of photons stored
in the atom is equal to the mean number of ingoing photons. In Section
\ref{master} we consider the case of a spherically symmetric atom stimulated by
monochromatic coherent light, we obtain the master equation which gives the
time evolution of the reduced atomic density matrix and we study the large-time
behavior of its solutions. In Section \ref{direct} we study the differential
(with respect to the angle) and total cross sections for the scattering of
laser light by the atom, as a function of the frequency of the stimulating
laser; in this section such cross sections are obtained from the direct
detection scheme. In Section \ref{heter}, starting from the balanced heterodyne
detection scheme, we obtain the power spectrum of the fluorescence light and
the elastic and inelastic cross sections. Section \ref{Dis} is devoted to a
discussion of the main features of the integral cross sections and of the power
spectrum.

\section{The model and the balance equation for the number of photons} \label{model}
First of all we want a model for an atom stimulated by a
laser (coherent light, not necessarily monochromatic); this means to choose as
initial state $\Psi(\xi, f)\in {\cal H}\otimes {\cal F}$ a generic state for
the atom and a coherent vector for the field \cite{Bar90}, i.e.
\begin{eqnarray}\label{2.1}
  \Psi(\xi,f) &=& \xi \otimes e(f)\,,
  \\ \nonumber
 \xi \in {\cal H}\,, \quad \|\xi\|&=&1\,,
  \quad f\in L^2({\mathbb R}_+;{\cal Z})\,.
\end{eqnarray}

Then, the atomic reduced statistical operator $\rho(t;\xi,f)$ is defined by the
partial trace over the Fock space
\begin{equation}\label{2.2}
  \rho(t;\xi,f) = {\mathrm Tr}_{{}_{\cal F}} \left\{ U(t) |\Psi(\xi,f) \rangle
  \langle \Psi(\xi,f)|  U(t)^\dagger  \right\}.
\end{equation}
Moreover, the quantity
\begin{equation}\label{2.3}
\langle N(t)\rangle_f = \langle U(t) \Psi(\xi,f) | N(t)\, U(t)
\Psi(\xi,f) \rangle
\end{equation}
represents the mean number of photons up to time $t$, after the interaction with
the atom, while
\begin{equation}\label{2.4}
\langle N(t)\rangle_f^0 =\langle \Psi(\xi,f) | N(t)\,
\Psi(\xi,f)\rangle = \int_0^t \|f(s)\|^2 {\mathrm d} s
\end{equation}
is the same quantity before such an interaction \cite{Bar90}; we can also say
that Eq.\ (\ref{2.4}) gives the mean number of ingoing photons entering the
system in the time interval $[0,t]$ and that Eq.\ (\ref{2.3}) gives the mean
number of outgoing photons leaving the system in the same time interval.

\end{multicols}
\vspace{-0.4cm}
\noindent\rule{0.49\textwidth}{0.4pt}\rule{0.4pt}{\baselineskip}
\widetext
\begin{proposition}\label{prop1}
The reduced statistical operator $\rho(t;\xi,f) $ satisfies the master equation
\begin{equation}\label{2.5}
\frac{{\mathrm d}\ }{ {\mathrm d} t} \, \rho(t;\xi,f) = {\cal L}\big(f(t)\big)
[\rho(t;\xi,f)]\,,
\end{equation}
where
\begin{mathletters}\label{2.6}
\begin{equation}\label{2.6a}
{\cal L}\big(f(t)\big)[\rho] = -{\mathrm i} \left[H\big(f(t)\big)\,,\,
\rho\right] + \frac 1 2 \sum_j \left( \left[ R_j \big(f(t)\big)\rho\,,\,
R_j\big(f(t)\big)^\dagger \right] +\left[ R_j\big(f(t)\big)\,,\, \rho
R_j\big(f(t)\big)^\dagger \right]\right),
\end{equation}
\begin{equation}\label{2.6b}
   R_j\big(f(t)\big) = R_j +\sum_i S_{ji} f_i(t)\,,
\end{equation}
\begin{equation} \label{2.6c}
H\big(f(t)\big) = H - \frac {\mathrm i} 2 \sum_{ij} \Bigl( R^\dagger_j S_{ji}
f_i(t) - \overline{f_i(t)} \, S_{ji}^{\ \dagger} R_j \Bigr);
\end{equation}
\end{mathletters}
moreover, we have
\begin{equation} \label{2.7}
\langle N(t) \rangle_f = \int_0^t {\mathrm Tr}_{{}_{\cal H}} \Bigl\{ \sum_j
R_j\big(f(s)\big)^{\dagger } R_j\big(f(s)\big) \rho(s;\xi,f) \Bigr\} {\mathrm
d} s\,.
\end{equation}
\end{proposition}
\noindent{}\hfill{
\rule[-\baselineskip]{0.4pt}{\baselineskip}\rule{0.49\textwidth}{0.4pt}}
\vspace{-0.4cm}
\begin{multicols}{2}
\narrowtext

\Proof By using the rules of QSC, it is possible to differentiate $\langle N(t)
\rangle_f$ and $\left\langle U(t) \Psi(\xi,f) | \,a \,U(t) \Psi(\xi,f)
\right\rangle$, where $a$ is a generic system operator. Then, one gets the
results by recalling that the increments of the field operators commute with
$U(t)$ and that ${\mathrm d} A_j(t) \Psi(\xi ,f)= f_j(t){\mathrm d} t\,
\Psi(\xi,f)$ and by using the definition of $\rho(t;\xi,f)$ given in Eq.\
(\ref{2.2}). \EndProof

In order to formulate physical requirements, let us start by considering the
case when no photon is injected into the system, i.e. $f=0$. In these
conditions it is natural to ask that the atom can emit at most one photon;
moreover, we require the existence of a unique equilibrium state which we
denote by $\rho_g$ (it will be the ground state). We take as canonical basis
$\{|+\rangle,\,|-\rangle\}$ in ${\cal H}$ an orthonormal basis which
diagonalises $\rho_g$, so that we can write
\begin{equation}\label{2.8}
\rho_g= pP_+ +(1-p)P_-
\end{equation}
for some $p$ in $[0,1]$, where $P_\pm$ are the orthogonal projections over the
vectors $|\pm\rangle$. We shall use also the Pauli matrices
\begin{mathletters}\label{2.9}
\begin{eqnarray}\label{2.9a}
\sigma_z = \left(\begin{array}{cc} 1& 0 \\ 0 &-1 \end{array}\right)&&, \qquad
\sigma_y = \left(\begin{array}{cc} 0& - {\mathrm i} \\ {\mathrm i}  & 0
\end{array}\right),
\\ \label{2.9b}
\sigma_+ = \left(\begin{array}{cc} 0& 1 \\ 0 &0 \end{array}\right)&&, \qquad
\sigma_- = \left(\begin{array}{cc} 0& 0 \\ 1 &0 \end{array}\right),
\end{eqnarray}
by which the two orthogonal projections $P_\pm$  can be written as
\begin{equation}\label{2.9c}
P_+ = \frac 1 2 (\openone + \sigma_z) = \sigma_+ \sigma_-\,, \quad P_- = \frac 1
2 (\openone - \sigma_z) = \sigma_- \sigma_+\,.
\end{equation}
\end{mathletters}

\begin{proposition}
We require
\begin{mathletters}\label{2.10}
\begin{eqnarray}\label{2.10a}
&& \langle N(t)\rangle_{f=0}  \leq 1\,, \qquad \forall \xi,\ \forall t,
\\ \label{2.10b}
&& \rho(t;\xi,0) \stackrel{t \to +\infty}{ \longrightarrow } \rho_g\,, \qquad
\forall \xi.
\end{eqnarray}
\end{mathletters}
\noindent Then, apart from an exchange of roles between the two states
$|+\rangle$ and $|-\rangle$, we obtain
\begin{equation}\label{2.11a}
\rho_g= P_-\,,
\end{equation}
\begin{mathletters}\label{2.11}
\begin{equation}\label{2.11b}
  H= \frac 1 2 \, \omega_0 \sigma_z\,, \qquad \omega_0 \in {\mathbb R}\,,
\end{equation}
\begin{equation}\label{2.11c}
  R_j = \langle e_j|\alpha \rangle\, \sigma_-\,, \qquad \alpha \in {\cal Z} \,,
  \ \alpha\neq 0\,.
\end{equation}
\end{mathletters}
\noindent Viceversa, Eqs.\ (\ref{2.11b}) and (\ref{2.11c}) imply Eqs.\
(\ref{2.10}) and (\ref{2.11a}).
\end{proposition}
\Proof By Eqs.\ (\ref{2.7}) and (\ref{2.10a}), $\langle N(t)\rangle_f$ is a
bounded and non decreasing function of $t$, so $\lim_{t\to +\infty}\langle
N(t)\rangle_f$ exists; then, Eqs.\ (\ref{2.7}) and (\ref{2.10b}) give
$\sum_j{\mathrm Tr}_{{}_{\cal H}} \left\{ R_j^{\dagger } R_j \rho_g
\right\}=0$. By the cyclic property of the trace and the positivity of $\rho_g$
and of $R_j \rho_g R_j^{\dagger }$, we get that this condition is equivalent to
$ R_j\, \rho_g= 0 $, $\forall j$.

Now, let us set $R_j = x_j\openone +y_j \sigma_z + z_j \sigma_+ + \alpha_j
\sigma_-$ (every operator on ${\mathbb C}^2$ can be written in this way). Then,
Eq.\ (\ref{2.8}) and $R_j\rho_g=0$ give $p(x_j+y_j)=0$, $(1-p)(x_j-y_j)=0$,
$(1-p)z_j=0$, $p\alpha_j=0$. For $p\in (0,1)$ this system of equations gives
$R_j=0$, which is not acceptable because in this case the equilibrium state is
not unique. For $p=0$ we get $x_j=y_j$ and $z_j=0$; we need also
\begin{equation}\label{2.12}
\sum_j|\alpha_j|^2\neq 0
\end{equation}
to have decay to an equilibrium state. We do not consider the case $p=1$,
because it is analogous to the previous one, apart from the exchange of
$|+\rangle$ and $|-\rangle$. Therefore we have Eq.\ (\ref{2.11a}) and
\begin{equation}\label{2.13}
R_j = \alpha_j \sigma_- + \beta_j P_+\,,
\end{equation}
with $\beta_j = 2 x_j$; by the convergence of $\sum_j R_j^\dagger R_j$, the
complex numbers $\alpha_j$ and $\beta_j$ can be seen as the components of two
vectors $\alpha$ and $\beta$ in $\cal Z$.

Eq.\ (\ref{2.5}) and (\ref{2.10b}) give ${\cal L}(0)[\rho_g]=0$; by Eqs.\
(\ref{2.6}), (\ref{2.11a}) and (\ref{2.13}) this condition reduces to
$[H,\rho_g]=0$. Because $H$ is selfadjoint and defined up to a constant, we
obtain Eq.\ (\ref{2.11b}).

Finally, let us choose $\xi=|+\rangle$. By using the relation $\sum_j
R_j^\dagger R_j = \left( \|\alpha\|^2 + \|\beta\|^2 \right) P_+$ and by
differentiating $\langle N(t) \rangle_{f=0}$ two times, we obtain
\[
\frac{{\mathrm d}^2\ }{{\mathrm d}t^2}\, \langle N(t)\rangle_{f=0} +
\|\alpha\|^2 \, \frac{{\mathrm d}\ }{{\mathrm d}t}\, \langle N(t)
\rangle_{f=0} =0\,,
\]
together with the initial conditions
\[
\langle N(0)\rangle_{f=0}=0\,,
\]
\[
 \frac{{\mathrm d}\ }{{\mathrm d}t}\,\langle N(0)\rangle_{f=0} =\|\alpha\|^2
 + \|\beta\|^2\,.
\]
This gives
\[
 \langle N(t)\rangle_{f=0} = \left( 1+ \frac{\|\beta\|^2}{\|\alpha\|^2} \right)
 \left(1- {\mathrm e}^{- \|\alpha\|^2t}\right);
 \]
then, condition (\ref{2.10a}) implies $\beta=0$ and Eqs.\ (\ref{2.12}) and
(\ref{2.13}) give  Eq.\ (\ref{2.11c}).

The last statement of the proposition follows by direct computations.
\EndProof

Now we have to find some physical restrictions on the possible forms of the
operator $S\in {\mathcal U}( {\mathcal H}\otimes {\mathcal Z})$. In
\cite{BarL98}, the case $f(t)= \lambda(t) \equiv \exp(-{\mathrm i}\omega
t)\theta(T-t)\, \lambda$ is considered, where $\theta(x)$ is the usual step
function and $\lambda \in {\mathcal Z}$; for $T\to +\infty$, $\lambda(t)$
represents a monochromatic coherent wave. Then, in \cite{BarL98} we asked
\begin{equation}
  \lim_{t\to +\infty} \lim_{T\to +\infty} \frac {\langle N(t) \rangle_\lambda}
  {\langle  N(t) \rangle_\lambda^0} =1\,,  \qquad \forall \lambda\in {\cal Z}\,,
  \quad \forall \omega\,,
\label{2.14}
\end{equation}
which is a form of flux conservation in the mean: if the possible physical processes
are absorption/emission of single photons and direct scattering without change of
atomic state, for large times the mean number of injected photons $\langle
N(t)\rangle_\lambda^0= \|\lambda\| t$ should be equal to the mean number of outgoing
photons $\langle N(t)\rangle_\lambda$.

The same restrictions on $S$ are obtained by requiring a balance equation on
the number of photons: the mean number of outgoing photons up to time $t$ plus
the mean number of photons stored in the atom must be equal to the mean number
of ingoing photons.

\begin{proposition}
Under assumptions (\ref{2.11}), the balance equation
\begin{eqnarray}\nonumber
  \langle N(t)\rangle_f &+&\frac 1 2\, {\mathrm Tr}_{{}_{\cal H}} \left\{ \sigma_z
  \bigl[ \rho(t;\xi,f) -  \rho(0;\xi,f) \bigr] \right\}
\\ &=& \langle N(t)\rangle_f^0 \label{2.15}
\end{eqnarray}
holds $\forall t$, $\forall \xi$, $\forall f$ if and only if one has
\begin{equation}\label{2.16}
  S =  P_+\otimes S^+ + P_-\otimes S^-\,, \qquad   S^\pm  \in {\cal U}
  ({\cal Z})\,.
\end{equation}
\end{proposition}

\Proof Any bounded operator on $\mathcal H \otimes Z$, like $S$, can always be
decomposed as
\begin{equation}\label{2.17}
  S=P_+\otimes S^+ + P_-\otimes S^- + \sigma_+ \otimes F^+ +\sigma_- \otimes
  F^-\,,
\end{equation}
where $ S^\pm,\, F^\pm$ are bounded linear operators on $\cal Z$; the unitarity
of $S$ implies some simple relations among $S^\pm$, $F^\pm$.

By using Eqs.\ (\ref{2.5}), (\ref{2.6}), (\ref{2.11}), we compute the time
derivative of ${\mathrm Tr}_{{}_{\cal H}} \{ \sigma_z \rho(t;\xi,f) \}$. Then,
we insert Eq.\ (\ref{2.6b}) into Eq.\ (\ref{2.7}) and, by using also Eq.\
(\ref{2.17}), we get
\begin{eqnarray*}
 \langle N(t)\rangle_f &-& \langle  N(t)\rangle_f^0
 \\
 &&{}+\frac 1 2\, {\mathrm Tr}_{{}_{\cal H}} \left\{
 \sigma_z \bigl[ \rho(t;\xi,f) -  \rho(0;\xi,f) \bigr] \right\}
\\
&=&\int_0^t {\mathrm d}s \, {\mathrm Tr}_{{}_{\cal H}} \Bigl\{ \Bigl[ \left\|
F^+ f(s) \right\|^2 P_- - \left\| F^- f(s) \right\|^2 P_+
\\
&&{}+ \left\langle S^+ f(s) \big| F^+ f(s) \right \rangle \sigma_+
\\
&& {}+ \left\langle F^+ f(s) \big| S^+ f(s) \right \rangle \sigma_- \Bigr]
\rho(s;\xi,f) \Bigr\}.
\end{eqnarray*}
By the arbitrariness of $t$, $f$ and $\xi$, condition (\ref{2.15}) is equivalent
to $F^\pm=0$ and Eq.\ (\ref{2.16}) is proved; the unitarity of $S^\pm$ follows
from the unitarity of $S$.
\EndProof

From now on we assume Eqs.\ (\ref{2.1}), (\ref{2.11}), (\ref{2.16}) to hold
and, always for physical reasons, we take
\begin{equation}\label{2.20}
  \omega_0 >0\,.
\end{equation}

In order to have an atom stimulated by a monochromatic coherent wave we take
\begin{equation}\label{3.1}
f(t)= \lambda(t) \equiv {\mathrm e}^{-{\mathrm i}\omega t}\theta(T-t)\,
\lambda\,, \quad \lambda \in {\mathcal Z}\,, \quad \omega>0\,.
\end{equation}
The step function $\theta$ is defined by $\theta(x)=1$ for $x\geq 0$ and
$\theta(x)=0$ for $x<0$, so that $\lambda(t)$ represents a monochromatic wave
for $T\to +\infty$.

Quantities like $\omega_0$, $\alpha$, $S^\pm$ are phenomenological parameters,
or, better, they have to be computed from some more fundamental theory, such as
some approximation to quantum electrodynamics. The whole model is meaningful
only for $\omega$ not too ``far'' from $\omega_0$ and $\omega_0$ must include
the Lamb shifts. In the final results one can admit a slight
$\omega$-dependence in the direct scattering matrices $S^\pm$.

\section{The master equation}\label{master}
In this section we study the master equation (\ref{2.5}) and the long time
behavior of the atom; the relations (\ref{2.6}), (\ref{2.11}), (\ref{2.16}),
(\ref{2.20}), (\ref{3.1}) hold.

\subsubsection*{The reduced statistical operator}
First of all, by setting
\begin{eqnarray}\nonumber
\rho_\lambda(t) &=& \lim_{T\to +\infty} \exp \left\{ {\textstyle \frac{\mathrm
i}{2}}\,  \sigma_z  (\beta+\omega t)\right\} \rho(t; \xi,\lambda)
\\  \label{3.2}
&&{}\times \exp \left\{ - {\textstyle\frac{\mathrm i}{2}}\, \sigma_z
(\beta+\omega t)\right\},
\\ \nonumber
\beta &=& \arg \left\{ - \langle S^- \lambda |\alpha\rangle\right\},
\end{eqnarray}
exactly as in Proposition \ref{prop1} we obtain the master equation
\begin{equation}\label{3.3}
\frac{{\mathrm d} \ }{{\mathrm d} t}\, \rho_\lambda(t) = {\cal L}_\lambda
\left[\rho_\lambda(t)\right]
\end{equation}
with the time independent Liouvillian
\begin{mathletters}\label{3.4}
\begin{eqnarray}\nonumber
{\cal L}_\lambda[\rho] &=& -{\mathrm i} [H_\lambda\,, \rho] +\frac 1 2 \sum_j
\left(\left[ R^\lambda_j \rho\,,\, R^{\lambda \dagger }_j \right]\right.
\\ \label{3.4a}
&&{}+ \left.\left[   R^\lambda_j\,,\, \rho R^{\lambda \dagger }_j
\right]\right),
\end{eqnarray}
\begin{eqnarray}\nonumber
R_j^\lambda &=& {\mathrm e}^{- {\mathrm i}\beta}\langle e_j | \alpha \rangle
\sigma_- + \langle e_j  |S^+ \lambda \rangle P_+
\\ &&{}+\langle e_j |S^- \lambda \rangle P_-\,, \label{3.4b}
\end{eqnarray}
\begin{equation}\label{3.4c}
H_\lambda = \frac 12\, (\omega_0 - \omega)\sigma_z - \frac {1}{ 2}\, |\langle
\alpha | S^-  \lambda \rangle |\sigma_y\,.
\end{equation}
\end{mathletters}

The general master equation for a two-level system is studied in \cite{Lendi};
in the following we shall use similar techniques, apart from a different
parametrization of the statistical operator, which turns out to be more
convenient in our case. By setting
\begin{mathletters}\label{3.5}
\begin{equation}\label{3.5a}
\rho_\lambda(t) = \left( \begin{array}{cc} u(t) &  v(t) \\  & \\
\overline{v(t)} &  1-u(t) \end{array} \right),
\end{equation}
\begin{equation} \label{3.5b}
\left\{\begin{array}{l} 0\leq u(t) \leq 1\,,  \\ \\ u(t)
\geq u^2(t) +  | v(t)|^2 \,,\end{array}\right.
\end{equation}
\end{mathletters}
\noindent where the conditions (\ref{3.5b}) express the fact that $\rho_\lambda
(t)$ is a statistical operator, we obtain from the master equation
\begin{equation}\label{3.6}
\frac{{\mathrm d} \ }{{\mathrm d} t}\, \bbox{u}(t) = - \bbox{G\, u }(t) +
\left(
\begin{array}{c} 0\\ \Omega /2  \\ \Omega /2
\end{array}\right),
\end{equation}
where
\begin{mathletters}
\begin{equation}\label{3.7b}
\bbox{u}(t) = \left( \begin{array}{c} u(t)\\ v(t)\\ \overline{v(t)}
\end{array} \right),
\end{equation}
\begin{equation}\label{3.7c}
\bbox{G}= \left( \begin{array}{ccc} \|\alpha\|^2 & -\Omega /2 & -\Omega /2
\\ & & \\
-{\mathrm e}^{{\mathrm i}\beta}\left \langle \alpha | \Delta S \lambda \right
\rangle +\Omega & b & 0
\\ & & \\
-{\mathrm e}^{-{\mathrm i}\beta}\left \langle \Delta S \lambda | \alpha \right
\rangle +\Omega & 0 &  \overline b \end{array} \right),
\end{equation}
\begin{eqnarray}\label{3.7a}
&&\Omega = 2\left|\left\langle \alpha | S^- \lambda \right \rangle \right|
\\ \label{3.7d}
&& b= \frac { \kappa^2} 2\, \|\alpha\|^2 - {\mathrm i} \left( \Delta \omega -
{\mathrm Im} \left \langle S^+ \lambda | P_\alpha  S^- \lambda \right\rangle
\right),
\\ \label{3.7e}
&&\kappa^2 = 1 + \|\Delta S \lambda\|^2\big/ \|\alpha\|^2
\\ \label{3.7g}
&&  \Delta S = S^+ -S^-\,,
\\ \label{3.7h}
&&\Delta \omega = \omega - \left( \omega_0 +  {\mathrm Im} \left\langle S^+
\lambda | P_\bot S^-  \lambda\right\rangle \right),
\\ \label{3.7i}
&&  \widetilde \alpha = \frac  \alpha{\|\alpha\|}\,, \qquad P_\alpha =
  \left| \widetilde \alpha  \right\rangle \left \langle \widetilde \alpha
  \right|,   \qquad P_\bot = \openone - P_\alpha\,.
\end{eqnarray}
\end{mathletters}
\noindent The quantity $\Omega$ can be interpreted as the bare Rabi frequency.
Moreover, we have
\begin{equation}\label{3.8}
  {\rm det}\, \bbox{G} =  \|\alpha\|^2 \left[ \left( \Delta \omega
  \right)^2 +  \Gamma^2/4 \right],
\end{equation}
with
\begin{eqnarray}\nonumber
\Gamma^2 &=&\kappa^4 \|\alpha\|^4+4 \kappa^2 \|\alpha\|^2\, {\mathrm Re}
\left\langle S^- \lambda | P_\alpha \left( S^+ + S^- \right) \lambda \right
\rangle
  \\ \nonumber
&&{}- 4 \left(  {\mathrm Im} \left\langle S^+ \lambda | P_\alpha S^- \lambda
\right \rangle \right)^2
  \\ \nonumber
&  \equiv &  \Bigl( \|\alpha\|^2 + \big\|P_\bot \Delta S \lambda\big\|^2+ 2
\left| \left\langle \widetilde \alpha | S^- \lambda \right \rangle \right |^2
\\  \nonumber
&&{}- 2\, {\mathrm Re} \left \langle S^+ \lambda | P_\alpha S^- \lambda \right
\rangle \Bigr)^2 + \left| \left \langle \widetilde \alpha \big| \left( S^+ +
S^- \right) \lambda \right\rangle \right|^2
\\
&&{}\times  \left[ \|\alpha\|^2\left(1+\kappa^2\right) + \big\|P_\bot \Delta S
\lambda\big\|^2 \right]. \label{3.9}
\end{eqnarray}
Let us note that $\|\alpha\|>0$ implies ${\rm det}\, \bbox{G} > 0$ and
$\Gamma^2 >0$.

\subsubsection*{The equilibrium state and the general solution of the master
equation}
The equilibrium state is given by
\begin{equation}\label{3.10}
  \lim_{t\to +\infty} \rho_\lambda(t) = \rho^\lambda_{\rm eq} =
  \left( \begin{array}{cc} u(\infty) & v(\infty) \\
  {}&{}\\
  \overline{v(\infty)} & 1-u(\infty) \end{array} \right),
  \end{equation}
where $u(\infty)$ and $v(\infty)$ are computed by equating to zero the time
derivative in Eq.\ (\ref{3.6}); then, we have $\bbox{u}(\infty)= \bbox{G}^{-1}
\bbox{w}$, which gives
\begin{mathletters}\label{3.11}
\begin{equation}\label{3.11a}
u(\infty) = \frac {\kappa^2 \Omega^2/4 }{ (\Delta \omega)^2 + \Gamma^2/4}\,,
\end{equation}
\begin{eqnarray}\label{3.11b}
v(\infty) &=&  \frac {\Omega/2} {(\Delta \omega)^2 + \Gamma^2/4}
  \\ \nonumber
&\times& \left( \frac{\kappa^2} 2 \, \|\alpha\|^2+ {\mathrm i} \Delta \omega
+{\mathrm i} \, {\mathrm Im}\, \langle S^+ \lambda | P_\alpha S^- \lambda
\rangle \right).
\end{eqnarray}
\end{mathletters}

For the computation of the fluorescence spectrum in Section \ref{heter}, we
shall have to solve the master equation (\ref{3.3}) also when the initial
condition is not a statistical operator. If
\[
\sigma = \left( \begin{array}{cc} \sigma_{11} & \sigma_{12}
\\ \sigma_{21} & \sigma_{22} \end{array} \right)
\]
is a generic $2\times 2$ matrix, we can always write
\begin{equation}\label{3.27}
{\mathrm e}^{{\mathcal L}_\lambda t}[\sigma] = (\sigma_{11} + \sigma_{22})
\rho_{\mathrm eq}^\lambda + \left(
\begin{array}{cc} d_1(t) & d_2(t) \\ d_3(t) & - d_1(t) \end{array} \right),
\end{equation}
where
\begin{equation}\label{3.28}
\bbox{d}(t) = {\mathrm e}^{-\bbox{G}t} \bbox{d}(0)\,, \qquad \bbox{d}(t) \equiv
\left( \begin{array}{c} d_1(t) \\ d_2(t) \\ d_3(t) \end{array} \right),
\end{equation}
\begin{eqnarray}
d_1(0)&=& \sigma_{11} - (\sigma_{11}+ \sigma_{22}) u(\infty)\,, \nonumber
\\ \label{3.29}
d_2(0)&=& \sigma_{12} - (\sigma_{11}+ \sigma_{22}) v(\infty)\,,
\\ \nonumber
d_3(0)&=& \sigma_{21} - (\sigma_{11}+ \sigma_{22}) \overline{ v(\infty)}\,.
\end{eqnarray}

\subsubsection*{Spherically symmetric atom stimulated by a collimated laser}
We end this section by particularizing our model to the case of a
\emph{spherically symmetric atom} stimulated by a \emph{well collimated laser}.

Let us recall that the Hilbert space $\mathcal Z$ contains the directions of
propagation of the electromagnetic field [see Eq.~(\ref{3.22})]. So, in order
to describe a laser beam propagating along the direction $\theta=0$, we have to
take
\begin{mathletters}\label{3.30}
\begin{equation}\label{3.30a}
  \lambda =\eta \|\alpha\|\,{\mathrm e}^{{\mathrm i} \delta} \widetilde \lambda\,, \qquad
  \eta>0\,,  \qquad  \delta\in [0,2\pi)  \,,
\end{equation}
\begin{equation}\label{3.30b}
  \widetilde \lambda (\theta,\phi) = \frac{ 1_{[0,\Delta \theta]}(\theta)}
  {\Delta \theta \sqrt{2\pi (1-\cos \Delta \theta)}} \,,
\end{equation}
\end{mathletters}
\noindent where $1_{[0,\Delta \theta]}(\theta)=1$ for $0\leq \theta \leq \Delta
\theta$, $1_{[0,\Delta \theta]}(\theta)=0$ elsewhere; in all the physical
quantities the limit $\Delta \theta \downarrow 0$ will be taken. Note that the
power of the laser $\hbar \omega\|\lambda\|^2=\hbar\omega \|\alpha\|^2 \eta^2/
{(\Delta\theta)^2}$ diverges for $\Delta \theta \downarrow 0$, because we need
a not vanishing atom-field interaction in the limit.

Let us denote by $Y_{lm}(\theta,\phi)$ the spherical harmonic functions; then,
the spherical symmetry of the atom requires
\begin{equation}\label{3.31}
  \widetilde \alpha(\theta,\phi) = Y_{00}(\theta,\phi) =1/\sqrt{4\pi}\,,
\end{equation}
\begin{equation}\label{3.32}
  S^\pm = \sum_{lm} {\mathrm e}^{2 {\mathrm i} \delta_l^\pm}\, |Y_{lm} \rangle
  \langle Y_{lm}|\,,
\end{equation}
where the quantities $\delta_l^+$ and $\delta_l^-$ are the phase shifts for the
direct scattering in the up and down atomic states respectively. Let us note
that we have
\begin{mathletters}\label{3.33}
\begin{equation}\label{3.33a}
\lim_{\Delta \theta \downarrow 0} \left\langle Y_{lm}\Big| \widetilde \lambda
\right \rangle = \delta_{m,0} \, \frac 1 2 \, \sqrt{2l+1}\,,
\end{equation}
\begin{equation} \label{3.33b}
\lim_{\Delta \theta \downarrow 0} \left\langle Y_{lm}\Big|
\left(S^\pm - \openone \right) \widetilde \lambda \right \rangle =
\delta_{m,0}\, {\mathrm i} \sqrt{2l+1}\, {\mathrm e}^{{\mathrm i} \delta_l^\pm}
\sin \delta_l^\pm\,.
\end{equation}
\end{mathletters}
\noindent Let us recall that $Y_{l0}(\theta,\phi)=\sqrt{\frac{2l+1}{4\pi}}\,
P_l(\cos \theta)$, where the functions $P_l(\xi)$ are the Legendre polynomials.

Now, we set
\begin{mathletters}\label{set}
\begin{eqnarray}\nonumber
  g_\pm(\theta) &&{}= \lim_{\Delta\theta \downarrow 0}\left(  \left( S^\pm -
  \openone \right) \widetilde \lambda \right)  (\theta,\phi) \\ \label{3.34}
&&{}= {\mathrm i} \sum_{l=0}^\infty \frac{2l+1}{\sqrt{4\pi}}\, {\mathrm
e}^{{\mathrm i} \delta^\pm_{l}} \sin \delta_l^\pm \, P_l(\cos \theta)\,,
\end{eqnarray}
\begin{eqnarray}\label{3.35}
&&\Delta g = g_+ - g_-\,,\qquad  s = \delta_0^+ - \delta_0^-\,,
\\ \label{5.28}
&&z= \frac {2\Delta \omega} {\|\alpha\|^2} \,, \qquad y=  z -
\frac{\eta^2}{2}\, \sin 2 s\,,
\\ \label{3.26}
&&\varepsilon= -\frac{\|\alpha\|^2} 4   \sum_{l=1}^\infty (2l+1) \sin
2(\delta_l^+ - \delta_l^-)\,,
\\ \nonumber
&&\zeta^2 = \left( 1+ \eta^2 \left\| P_\bot \Delta g \right\|^2 \right)^2
\\ &&\quad \
{}+\eta^2 \left( 1+ \kappa^2 + \eta^2 \left\| P_\bot \Delta g \right\|^2
\right), \label{3.36m}
\\ \label{set5}
&&b^\prime= \kappa^2 - {\mathrm i} \left( z + \frac{\eta^2} 2\, \sin 2s
\right),
\\ \label{set6}
&&\bbox{G}^\prime= \left( \begin{array}{ccc} 2 & - \eta & - \eta
 \\ & & \\
2\eta {\mathrm e}^{{\mathrm i}s} \cos s  &
 b^\prime & 0 \\ & & \\
2\eta {\mathrm e}^{-{\mathrm i}s} \cos s & 0
 &  \overline{ b^\prime} \end{array} \right).
\end{eqnarray}
\end{mathletters}
\noindent In order that all these quantities be finite, we require also
\begin{mathletters}\label{3.37}
\begin{eqnarray}\label{3.37a}
&&\sum_{l=0}^\infty (2l+1) \sin^2 \delta_l^\pm < +\infty\,,
\\ \label{3.37b}
&&\sum_{l=0}^\infty (2l+1) \left| \sin 2 \left(\delta_l^+ -\delta_l^-\right)
\right|< +\infty\,.
\end{eqnarray}
\end{mathletters}
\noindent Then, we have
\begin{mathletters}\label{3.36}
\begin{eqnarray} \label{3.36k}
&&\beta= \pi - \delta - 2\delta_0^-\,,
\\ \label{3.36b}
&&b= \frac {\|\alpha\|^2} 2 \, b^\prime\,,
\\  \label{3.36a}
&&\bbox{G}= \frac{ \|\alpha\|^2 } 2\, \bbox{G}^\prime\,,
\\ \label{3.36g}
&&\Delta \omega = \omega - \left(\omega_0 + \eta^2 \varepsilon\right),
\\ \label{3.36y}
&& \Omega= \eta \|\alpha\|^2\,,
\\ \label{3.36h}
&&\Gamma^2 = \zeta^2\|\alpha\|^4\,,
\\ \label{3.36c}
&&\kappa^2 = 1 + \eta^2\|\Delta g\|^2\,,
\\  \label{3.36i}
&&u(\infty) = \frac{ \eta^2\kappa^2 }{z^2+\zeta^2}\,,
\\ \label{3.36j}
&&v(\infty) = \frac{ \eta} {z^2+\zeta^2} \left( \kappa^2 +{\mathrm i} y
\right),
\\ \label{3.36l} &&\Delta g(\theta) =  {\mathrm i}\, \frac{{\mathrm
e}^{{\mathrm i }\left(\delta_0^++ \delta_0^-\right)}} {\sqrt{4 \pi}} \, \sin
s+\big( P_\bot\Delta g\big)(\theta)\,,
\\ \nonumber
&&\big( P_\bot\Delta g\big)(\theta)={\mathrm i} \sum_{l=1}^\infty \frac {2l+1}{
\sqrt{4\pi}}\, {\mathrm e}^{{\mathrm i}\left(\delta_l^+ + \delta_l^- \right)}
\\ \label{3.36d}
&&{}\qquad\qquad\qquad{}\times \sin \left(\delta_l^+ - \delta_l^- \right)
P_l(\cos \theta)\,,
\\ \label{3.36e}
&&\|\Delta g\|^2 =  \sin^2 s + \|P_\bot\Delta g\|^2
\\ \label{3.36f}
&&\|P_\bot\Delta g\|^2 = \sum_{l=1}^\infty (2l+1) \sin^2\left(\delta_l^+ -
\delta_l^- \right),
\\ \label{3.36x}
&&\|g_\pm\|^2 = \sum_{l=0}^\infty (2l+1) \sin^2 \delta_l^\pm\,,
\end{eqnarray}
\end{mathletters}

\paragraph*{Low intensity laser.}

For future use, it is useful to particularize the previous quantities to the
case of a laser of vanishing intensity, i.e.\ $\eta=0$:
\begin{mathletters}\label{3.23}
\begin{eqnarray}\label{3.23a}
&&b^\prime= 1 - {\mathrm i} z\,, \qquad \kappa^2=1\,, \qquad \zeta=1\,,
\\ \label{3.23b}
&&\Omega=0\,, \qquad \Gamma=\|\alpha\|^2\,, \qquad \Delta \omega = \omega -
\omega_0\,.
\end{eqnarray}
\end{mathletters}

\paragraph*{No direct scattering.}

The usual model of a two-level atom, with only absorption/emission and no
direct scattering, is characterized by $S^\pm = \openone$, so that the previous
quantities reduce to
\begin{mathletters}\label{3.38}
\begin{eqnarray}\label{3.38a}
&& g_\pm = 0\,, \qquad s=0\,, \qquad y=z\,,
\\ \label{3.38b}
&& \beta = \pi - \delta\,, \qquad \kappa^2=1\,, \qquad \zeta= \sqrt{1+2\eta^2}
\\ \label{3.38c}
&& \Gamma^2=\|\alpha\|^4+ 2 \Omega^2\,, \qquad \Delta \omega = \omega -
\omega_0\,,
\\ \label{3.38d}
&& u(\infty) = \frac{ \eta^2}{ z^2+\zeta^2}\,,
\\ \label{3.38e}
&& v(\infty) = \frac{ \eta} {z^2+\zeta^2} \left( 1 +{\mathrm i} z \right),
\\ \label{3.38f}
&&\bbox{G}^\prime= \left( \begin{array}{ccc} 2 & - \eta & - \eta
 \\ & & \\
2\eta   & 1 - {\mathrm i}z & 0
\\ & &
\\ 2 \eta  & 0
 &  1 + {\mathrm i} z  \end{array} \right).
\end{eqnarray}
\end{mathletters}

\section{Direct detection and total cross section}\label{direct}
By direct detection, it is possible to measure the intensity of the light (or
to count the photons) propagating in a small solid angle $\Delta \Upsilon$
around some direction, which we take different from the direction $\theta=0$ of
the incoming beem. The observable ``number of photons in $\Delta\Upsilon$ up to
time $t$'' is represented by\begin{equation}\label{4.1}
  N(t;\Delta\Upsilon) = \sum_{i,j} \langle e_i| 1_{\Delta\Upsilon} \,e_j \rangle\,
  \Lambda_{ij}(t)\,,
\end{equation}
where $ 1_{\Delta\Upsilon}(\theta,\phi)=1$ for $(\theta,\phi)\in \Delta
\Upsilon$ and $ 1_{\Delta\Upsilon}(\theta,\phi)=0$ elsewhere. The fact that the
direction of detection is different from the beam direction is expressed by
\begin{equation}\label{4.2}
1_{\Delta\Upsilon}\, \lambda=0\,.
\end{equation}
Then, the mean number of photons up to time $t$ per unit of solid angle around
$(\theta, \phi)$ is given by
\begin{equation}\label{4.3}
  \langle n(\theta,\phi;t)\rangle = \frac{1}{|\Delta\Upsilon|}
  \langle  U(t) \Psi(\xi,\lambda)
  |  N(t;\Delta\Upsilon) U(t) \Psi(\xi,\lambda)\rangle\,,
\end{equation}
where $\Psi$, $\lambda(t)$, $\lambda$ are given by Eqs.\ (\ref{2.1}),
(\ref{3.1}), (\ref{3.30}) and $|\Delta\Upsilon|= \int\!\!\int_{\Delta \Upsilon}
\sin \theta \,{\mathrm d}\theta {\mathrm d} \phi$; the limits $T\to +\infty $,
$\Delta \theta \downarrow 0$, $\Delta \Upsilon \downarrow \{(\theta, \phi)\}$
are understood.

The (angular) differential cross section is proportional to the outgoing flux
per unit of solid angle $ \langle n(\theta,\phi;t)\rangle/t$ divided by the
incoming flux $\langle \Psi(\xi,\lambda)|N(t) \Psi(\xi,\lambda)\rangle /t$; so
we have
\begin{eqnarray}\nonumber
  \sigma(\theta,\phi) &=& A_0 \lim_{t\to +\infty} \frac {\langle n(\theta,\phi;t)
  \rangle} {\langle \Psi(\xi,\lambda) | N(t) \Psi(\xi,\lambda) \rangle}
  \\ \label{4.4}
  &=& \frac {A_0}{ \|\lambda\|^2}
  \, \lim_{t\to +\infty} \frac 1 t \, \langle n(\theta,\phi;t) \rangle\,,
\end{eqnarray}
where $A_0$ is a kinematical factor to be determined and with dimensions of an
area. To determine $A_0$ let us consider the cross section for direct photon
scattering by the up or down atomic state, for which the Bohr-Peierls-Placzek
formula (or optical theorem) gives $\sigma(\theta,\phi) = |q(\theta)|^2$,
$\sigma_{{}_{\rm TOT}}= 2 \, \frac {2\pi c } \omega\,  {\mathrm Im}\, q(0)$;
the total cross section is the integral of the differential one on the whole
solid angle. In our case we have to take $\alpha=0$ and from Eq.\ (\ref{4.4})
we get $\sigma(\theta,\phi)=A_0\left| \left( \left( S^\pm-\openone \right)
\lambda \right) (\theta,\phi)\right|^2/ \|\lambda\|^2$ and, by the unitarity of
$S^\pm$,
\begin{eqnarray*}
\sigma_{{}_{\rm TOT}}&=& \frac{A_0}{\|\lambda\|^2} \left\| \left( S^\pm
-\openone \right)\lambda\right\|^2
\\ &=&-\frac{2A_0\sqrt\pi\,
\Delta\theta}{\|\lambda\|} \,{\mathrm Im}\, {\mathrm i}{\mathrm e}^{-{\mathrm
i}\delta}
 \left( \left( S^\pm -\openone \right)\lambda \right)(0,0)\,.
\end{eqnarray*}
Then, we must have $q(\theta)= -{\mathrm i} \sqrt{A_0}\, \Delta \theta \, g_\pm
(\theta) $ and, by imposing the optical theorem, we get $A_0= \left(\frac{2\pi
c}{\omega}\right)^2\frac1 {\pi(\Delta\theta)^2}$. Up to now we have not taken
into account the polarization degrees of freedom. If they are taken into
account and the cross section for not polarized light is considered, a $3/2$
extra-factor is obtained $(\!\!{}$\cite{CT92} pp.\ 532--533) and Eq.\
(\ref{4.4}) becomes
\begin{equation}\label{4.5}
\sigma(\theta,\phi) = \left(\frac{2\pi c}{\omega}\right)^2\frac 3
{2\pi\eta^2\|\alpha\|^2} \, \lim_{t\to +\infty} \frac 1 t \, \langle
n(\theta,\phi;t) \rangle\,.
\end{equation}

To compute $\sigma(\theta,\phi)$ we differentiate Eq.\ (\ref{4.3}) by using the
rules of QSC, we use Eq.\ (\ref{4.2}) and then we apply the transformation
(\ref{3.2}); the final result is
\begin{equation}\label{4.6}
  \frac{{\mathrm d} \ }{{\mathrm d} t} \left\langle n(\theta,\phi;t)\right\rangle =
  {\mathrm Tr} \left\{ R(\theta,\phi)^\dagger
  \,R(\theta,\phi)\,  \rho_\lambda (t) \right\},
\end{equation}
\begin{equation}\label{4.7}
R(\theta,\phi)= {\mathrm e}^{{-\mathrm i}\beta} \, \frac{\|\alpha\|}
{\sqrt{4\pi}} \,\sigma_- + {\mathrm e}^{{\mathrm i}\delta}\eta \|\alpha\|\bigl[
g_+(\theta) P_+  +   g_-(\theta) P_-\bigr].
\end{equation}
Then, Eq.\ (\ref{4.5}) gives
\begin{equation}\label{4.8}
\sigma(\theta,\phi) = \frac{6\pi c^2}{\eta^2\|\alpha\|^2\omega^2} \, {\mathrm
Tr} \left\{ R(\theta,\phi)^\dagger \,R(\theta,\phi)\, \rho^\lambda_{\rm eq}
\right\}.
\end{equation}
Finally, by computing the trace and by using the results of the previous
section we obtain the differential cross section and, by integrating it, the
total one:
\begin{eqnarray}\nonumber
\sigma(\theta,\phi) &=&  \frac{6\pi c^2}{\omega^2}  \biggl\{ |g_-(\theta)|^2 +
\frac{\kappa^2}{z^2+\zeta^2}
\\  \label{4.9}&\times&
\left[ \frac{1}{ 4\pi} + \eta^2 \left( | g_+(\theta)|^2 -
|g_-(\theta)|^2\right)\right]
  \\ \nonumber &-&
\frac{2}{\sqrt{4\pi}\left (z^2+\zeta^2\right)}\, {\mathrm Re} \left[{\mathrm
e}^{-2{\mathrm i} \delta_0^-} g_-(\theta) \left(\kappa^2 - {\mathrm i} y
\right) \right] \biggr\},
\end{eqnarray}
\begin{eqnarray}\nonumber
\sigma_{{}_{\rm TOT}} &=&  \frac{6\pi c^2}{\omega^2} \biggl\{ \| g_-\|^2
+\frac{\kappa^2}{ z^2+ \zeta^2}
\\  \label{4.10} &\times&
\left[1 + \eta^2\left(\|g_+\|^2 - \|g_-\|^2 \right)\right]
 \\ \nonumber
&-&\frac{1}{ z^2+\zeta^2}\left(y \sin 2\delta_0^- + 2 \kappa^2 \sin^2
\delta_0^- \right) \biggr \}.
\end{eqnarray}

Let us note that the angular dependence in $\sigma(\theta,\phi)$ is entirely
due to $g_\pm(\theta)$ (\ref{3.34}) and, so, to the presence of the
$\Lambda$-term in Eq.\ (\ref{1.2}).

By some algebraic manipulations $\sigma_{{}_{\rm TOT}} $ can be rewritten in a
more perspicuous form:
\begin{eqnarray}\nonumber
\frac {\omega^2}{ 6\pi c^2}\, \sigma_{{}_{\rm TOT}} &=& \frac{\left(z\sin
\delta_0^- - \cos  \delta_0^- \right)^2 + \eta^2 A}{z^2 +\zeta^2}
\\ &&{}+
\left\| P_\bot g_- \right\|^2 \, \frac{z^2+ B}{z^2 +\zeta^2}\,,\label{4.14}
\end{eqnarray}
\begin{eqnarray}\nonumber
A&=& \sin^2 \delta_0^+ + \kappa^2 \|g_+\|^2 + \left\| P_\bot \Delta g
\right\|^2
\\ \label{4.12}
&&{}\times \left[ 1+ \eta^2 \left( 1+ \left\| P_\bot \Delta g \right\|^2
\right) \sin^2 \delta_0^- \right],
\end{eqnarray}
\begin{equation}
B= \left( 1+ \eta^2 +\eta^2\left\| P_\bot \Delta g \right\|^2  \right)
\left(1+\eta^2 \left\| P_\bot \Delta g \right\|^2 \right). \label{4.13}
\end{equation}
According to the values of the various coefficients different line shapes
appear, which are known as Fano profiles $(\!\!{}$\nolinebreak\cite{CT92} pp.\
61--63). These shapes are typical of the interference among various channels,
when one of them has an amplitude with a pole near the real axis in the complex
energy plane (see also Eq.\ (\ref{4.9bis}) below); in our case the channels are
direct scattering in the up state, direct scattering in the down state and
fluorescence. Some plots of $\frac{\omega^2}{6\pi c^2}\, \sigma_{{}_{\rm TOT}}$
are given in Fig.\ \ref{fig1}; the independent variable is the ``reduced"
detuning $\widetilde z= (\omega -\omega_0)/\|\alpha\|^2$, the other parameters
are given in the caption of Fig.\ \ref{fig1}; the same figure contains plots of
elastic and inelastic cross sections, which will be discussed in Sections
\ref{heter} and \ref{Dis}.

Whichever the line shape be, there is a strong variation of the cross section
for $\omega$ around $\omega_0 +\eta^2\varepsilon$ [see Eqs.\ (\ref{5.28}),
(\ref{3.26}), (\ref{3.36g})]. The intensity dependent shift $\eta^2\varepsilon$
of the resonance frequency has received various names in the literature; a very
suggestive one is \emph{lamp shift}, a name suggested by A.\ Kastler in
\cite{Kast}. Note that in our two-level system the lamp shift is not vanishing
only if the two states respond differently to direct scattering; moreover, only
the contributions different from the $s$-wave ones do matter. Let us stress
that also the width $\Gamma$ of the resonance and the whole line shape are
intensity dependent.

\paragraph*{No direct scattering.}
Let us also note that when the direct scattering is negligible, i.e.\ when
Eqs.\ (\ref{3.38}) hold, Eq.\ (\ref{4.10}) reduces to
\begin{equation}\label{4.11}
   \sigma_{{}_{\rm TOT}} = \frac{6\pi c^2}{ \omega^2}
 \, \frac {\|\alpha\|^4/4}{(\Delta \omega)^2 + \Gamma^2/4}\,.
\end{equation}
For a laser with negligible intensity, i.e.\ when $\eta \downarrow 0$, Eq.\
(\ref{4.11}) reduces to the cross section for resonant scattering, given in
\cite{CT92} pp.\ 530--533; for $\eta \neq 0$, we have a power broadening (see
Eq.\ (\ref{3.38c})) of the resonance line, which maintains a Lorentzian shape
\cite{Eze1}.

By comparing the general case (\ref{4.14}) with the usual one (\ref{4.11}), we
see that the main differences are that in the general case we have lamp shift,
asymmetric line shape and bigger power broadening.

\paragraph*{Low intensity laser.}

For $\eta=0$ Eqs.\ (\ref{3.23}) hold and Eqs.\ (\ref{4.9}) and (\ref{4.14})
reduce to
\begin{equation}\label{4.9bis}
\frac {\omega^2} {6\pi c^2}\, \sigma(\theta,\phi) = \left| g_-(\theta) - \frac
{{\mathrm i\, e}^{2{\mathrm i} \delta^-_0}} {\sqrt{4\pi} \left( z+ {\mathrm i}
\right)} \right|^2,
\end{equation}
\begin{equation}\label{4.15}
\frac {\omega^2} {6\pi c^2}\, \sigma_{{}_{\rm TOT}} =  \|P_\bot g_-\|^2 +
\frac{\left(z \sin \delta_0^- - \cos \delta_0^-\right)^2} {z^2 +1}\,.
\end{equation}

\section{Heterodyne detection}\label{heter}
\subsection{Power spectrum}\label{spectrum}
The best way to obtain the spectrum of our stimulated atom is by means of the
\emph{balanced heterodyne detection} scheme; the output current of the detector
is represented by the operator \cite{Bar90,Goslar}
\begin{equation}\label{5.1}
I(\nu,h;t) = \int_0^t F(t-s) j(\nu,h;{\mathrm d} s)\,,
\end{equation}
where $F(t)$ is the detector response function, say
\begin{equation}\label{5.2}
F(t) = k_1 \sqrt{\frac{\gamma}{4\pi}}\, \exp\left( -\frac{\gamma}{2}\, t
\right), \qquad \gamma>0\,,
\end{equation}
$k_1\neq 0$ has the dimensions of a current, $j$ is essentially a field
quadrature
\begin{equation}\label{5.3}
j(\nu,h;{\mathrm d} s) = \overline{ q}\, {\mathrm e}^{{\mathrm i} \nu s} \,
{\mathrm d} A_h(s) + \text{h.c.}\,,
\end{equation}
\begin{equation}\label{5.9}
{\mathrm d} A_h(t) = \sum_j \langle h|e_j\rangle\, {\mathrm d}A_j(t)\,,
\end{equation}
$q$ is a phase factor, $q\in \mathbb C$, $|q|=1$, $\nu$ is the frequency of the
local oscillator and $h\in \mathcal Z$, $\|h\|=1$; $h$ contains information on
the localization of the detector, say
\begin{equation}\label{5.4}
h(\theta^\prime, \phi^\prime) = \frac{1}{\sqrt{|\Delta \Upsilon|}}\, 1_{\Delta
\Upsilon} (\theta^\prime,\phi^\prime)\,,
\end{equation}
where $\Delta \Upsilon$ is again the small solid angle around $(\theta,\phi)$
introduced in the previous section. From the canonical commutation relations
for the fields one has
\begin{eqnarray}\nonumber
&&\big[I(\nu_1, h_1; t_1), I(\nu_2, h_2; t_2)] = \int_0^{\min\{t_1,t_2\}}
{\mathrm d}s \, F(t_1-s)
\\
&& \qquad{}\times  F(t_2-s) \left( {\mathrm e}^{{\mathrm i}(\nu_1 - \nu_2) s}
\langle h_1| h_2 \rangle - \text{c.c.} \right) ; \label{5.6}
\end{eqnarray}
so, $I(\nu_1, h_1; t_1)$ and $I(\nu_2, h_2; t_2)$ are compatible observables
for any choice of the times either if $\nu_1=\nu_2$ and $h_1=h_2$ either if
$\langle h_1|h_2\rangle=0$. Under the same conditions also the $j$'s commute.

In the following for the quantum expectation of any operator $B$ we shall use
the notation
\begin{equation}\label{5.11}
\langle B\rangle_\lambda^T = \langle U(T) \Psi(\xi,\lambda)| B U(T)
\Psi(\xi,\lambda) \rangle\,.
\end{equation}

In the long run the output mean power is given by
\begin{equation}\label{5.5}
P(\nu,h) = \lim_{T \to +\infty} \frac{k_2}{T} \int_0^T \big\langle \big(
I(\nu,h;t)\big)^2 \big\rangle_\lambda^T \,{\mathrm d}t\,;
\end{equation}
$k_2>0$ has the dimensions of a resistance, it is independent of $\nu$, but it
can depend on the other features of the detection apparatus. In this section
$\lambda (t)$ is given by Eq.\ (\ref{3.1}); the limit case (\ref{3.30}) will be
considered in the next one. As a function of $\nu$, $P(\nu,h)$ gives the
\emph{power spectrum} observed in the ``channel $h$''; in the case of the
choice (\ref{5.4}) it is the spectrum observed around the direction
$(\theta,\phi)$. Proposition \ref{prop4} relates $P(\nu,h)$ to normal ordered
quantum expectations of products of field operators and gives a sum rule which
relates $P(\nu,h)$ to $\|\lambda\|^2$; let us note that $\hbar \omega
\|\lambda\|^2$ is the total power of the input monochromatic state $\lambda(t)$
(\ref{3.1}). Proposition \ref{prop5} identifies an elastic and an inelastic
contribution to the power and reduces the computation of $P(\nu,h)$ to the
solution of the master equation (\ref{3.4}). For the use of QSC in the
computation of the spectrum of a two-level atom see also Ref.\ \cite{Maass}.

\begin{proposition}\label{prop4}
The mean power $P(\nu,h)$ can be expressed as
\begin{eqnarray}\label{5.8}
P&&(\nu,h) = \frac k{4\pi} + \lim_{T\to +\infty} \frac{k}{ 2\pi T}
\\
&&{}\times\left\{ \Big\langle \int_0^T {\mathrm d} A^\dagger_h(t) \int_0^t
{\mathrm d} A_h(s) \, {\mathrm e}^{ - \left( \frac \gamma 2 +{\mathrm i} \nu
\right) (t-s) } \Big\rangle_\lambda^T + \text{\rm c.c.} \right\},\nonumber
\end{eqnarray}
where $k= k_1^{\,2} k_2$; Eq.\ (\ref{5.8}) holds almost everywhere in $\nu$.

We have also
\begin{equation}\label{5.21}
\int_{-\infty}^{+\infty} \left[ P(\nu,h) - \frac{k}{4\pi} \right]{\mathrm d}\nu
= \lim_{T\to +\infty} \frac k T \, \langle \Lambda_{hh}(T) \rangle_\lambda^T\,,
\end{equation}
where $\Lambda_{hh}(T) = \sum_{ij} \langle e_i|h\rangle \Lambda_{ij} \langle h|
e_j\rangle $; moreover, for any c.o.n.s.\ $\{h_j\}$ in $\mathcal Z$, the
following sum rule holds:
\begin{equation}\label{5.10}
\sum_j \int_{-\infty}^{+\infty} \left[ P(\nu,h_j) - \frac{k}{4\pi}
\right]{\mathrm d}\nu = k \|\lambda\|^2\,.
\end{equation}
\end{proposition}

\Proof By inserting Eqs.\ (\ref{5.1}) and (\ref{5.2}) into the definition
(\ref{5.5}) and by changing order of integration, one gets
\begin{eqnarray*}
P(\nu,h) &=& \lim_{T\to +\infty} \frac{k}{ 4\pi T} \int_0^T \int_0^T
\left({\mathrm e}^{- \frac \gamma 2 \left| t-s\right|} - {\mathrm e}^{- \gamma
\left( T- \frac{t+s} 2 \right)} \right)
\\
&&{}\times  \langle j(\nu,h;{\mathrm d}t) j(\nu,h;{\mathrm d}s)
\rangle_\lambda^T\, .
\end{eqnarray*}
The term containing the factor $\exp \left[ - \gamma \left( T- \frac{t+s} 2
\right)\right]$ vanishes for $T\to +\infty$ and one obtains
\begin{eqnarray}\nonumber
P(\nu,h) &=& \lim_{T\to +\infty} \frac{k}{ 4\pi T} \int_0^T \int_0^T {\mathrm
e}^{- \frac \gamma 2 \left| t-s\right|}
\\
&&{}\times \langle j(\nu,h;{\mathrm d}t) j(\nu,h;{\mathrm d}s)
\rangle_\lambda^T \,.\label{5.7}
\end{eqnarray}

By using the canonical commutation relations and normal ordering, we have
\begin{eqnarray*}
P(\nu,h) - \frac k {4\pi}&=& \lim_{T\to +\infty} \frac{k}{ 2\pi T} \int_{t\in
(0,T)} \int_{s\in (0,t)} {\mathrm e}^{- \frac \gamma 2 \left( t-s\right)}
\\
&&{}\times  \langle\, :j(\nu,h;{\mathrm d}t) j(\nu,h;{\mathrm
d}s):\,\rangle_\lambda^T
\\
&=& \lim_{T\to +\infty} \frac{k}{ 2\pi T} \int_{t\in (0,T)} \int_{s\in (0,t)}
{\mathrm e}^{- \frac \gamma 2 \left( t-s\right)}
\\
&&{}\times \Big\langle \Bigl\{{\mathrm e}^{- {\mathrm i} \nu (t-s)}\, {\mathrm
d} A_h^\dagger(t) {\mathrm d}A_h(s)
\\
&&{}+ \overline{q}^2{\mathrm e}^{ {\mathrm i} \nu (t+s)}\, {\mathrm d} A_h(t)
{\mathrm d}A_h(s)\Bigr\} \Big\rangle_\lambda^T + \text{c.c.}
\end{eqnarray*}
The factor $\exp[{\mathrm i} \nu(t+s)]$, when integrated over $\nu$ from
$\nu_1$ to $\nu_2$, gives rise to $\big\{ \exp[{\mathrm i} \nu_2(t+s)] -
\exp[{\mathrm i} \nu_1(t+s)]\big\}\big/ \{{\mathrm i} (t+s)\}$, which is not
singular for $t>0$ and $s>0$; then, the integral containing this factor
vanishes for $T\to +\infty$ and Eq.\ (\ref{5.8}) is proved.

Now let us observe that
\[
\int_0^T{\mathrm d} A^\dagger_h(t)\int_0^T {\mathrm d} A_h(s)\, \delta(t-s) =
\Lambda_{hh}(T) \,.
\]
By integrating over $\nu$ the second term in the r.h.s.\ of Eq.\ (\ref{5.8}) a
Dirac delta comes out and by adding the complex conjugated term a double
integral for $s\in (0,T)$ and $t\in (0,T)$ is obtained; then, by the previous
observation Eq.\ (\ref{5.21}) is obtained. By Eqs.\ (\ref{1.1}), (\ref{2.3}),
(\ref{5.21}), we obtain
\[
\sum_j \int_{-\infty}^{+\infty} \left[ P(\nu,h_j) - \frac{k}{4\pi} \right]
{\mathrm d}\nu= \lim_{T\to +\infty} \frac{k}{T} \langle N(T)\rangle_\lambda^T
\,.
\]
Finally, by Eqs.\ (\ref{2.15}), (\ref{2.4}), (\ref{3.1}), the sum rule
(\ref{5.10}) is obtained. \EndProof

\begin{proposition}\label{prop5}
The mean power can be decomposed as the sum of three positive contributions
\begin{equation}\label{5.13}
P(\nu,h) = \frac k {4\pi} + P_{\mathrm el}(\nu,h) + P_{\mathrm inel}(\nu,h)\,,
\end{equation}
where
\begin{equation}\label{5.17}
P_{\mathrm el}(\nu,h) = k\left| r(h) \right|^2 \frac 1 \pi\, \frac {\gamma/2}{
(\nu - \omega)^2 + \gamma^2/4}\,,
\end{equation}
\begin{eqnarray}\nonumber
P_{\mathrm inel}(\nu,h) &=& \frac k {2\pi} \int_0^{+\infty} {\mathrm d}t\, \exp
\left[ - \left( \frac\gamma 2 + {\mathrm i} (\nu - \omega) \right) t \right]
\\
&\times & {\mathrm Tr} \left\{ D(h)^\dagger \left( {\mathrm e}^{ {\mathcal
L}_\lambda t } \left[ D(h) \rho_{\mathrm eq}^\lambda \right]\right) \right\} +
\text{\rm c.c.}, \label{5.16}
\end{eqnarray}
\begin{mathletters}\label{5.18}
\begin{eqnarray}\label{5.18a}
D(h) &=& R(h) - r(h)\,,
\\ \label{5.18b}
r(h) &=& {\mathrm Tr} \left\{ R(h) \rho_{\mathrm eq}^\lambda \right\},
\\ \nonumber
R(h) &=& {\mathrm e}^{- {\mathrm i}\beta}\langle h|\alpha \rangle \sigma_- +
\langle h| S^+\lambda \rangle P_+ + \langle h| S^- \lambda\rangle P_-
\\
&=& \sum_j \langle h|e_j \rangle R_j^\lambda\,. \label{5.18c}
\end{eqnarray}
\end{mathletters}
\end{proposition}

\Proof Let us start from Eq.\ (\ref{5.8}). We can write
\begin{eqnarray*}
\langle {\mathrm d} A_h^\dagger(t) {\mathrm d} A_h(s) \rangle_\lambda^T &=&
\langle \Psi(\xi,\lambda)| U(T)^\dagger {\mathrm d} A_h^\dagger(t) U(T)
\\
&\times& U(T)^\dagger  {\mathrm d} A_h(s) U(T) \Psi(\xi,\lambda)\rangle
\end{eqnarray*}
with $T>t>s$. By the rules of QSC (see the ``output fields'' in \cite{Bar90},
Section 3), we obtain
\begin{eqnarray*}
U(T)^\dagger &&{\mathrm d} A_h(t) U(T) = U(t)^\dagger \bigl\{ \langle h|\alpha
\rangle \sigma_- \, {\mathrm d}t
\\
{}+&& \sum_j\bigl( \langle h|S^+e_j \rangle P_+ +  \langle h|S^- e_j \rangle
P_- \bigr) {\mathrm d}A_j(t) \bigr\} U(t)\,.
\end{eqnarray*}
By using this result we can write
\begin{eqnarray}\label{bb}
\Big\langle &&\int_0^T {\mathrm d} A_h^\dagger(t)\int_0^t {\mathrm d} A_h(s) \,
{\mathrm e}^{-\left( \frac \gamma 2 + {\mathrm i} \nu \right)(t-s)} \Big
\rangle_\lambda^T
\\ \nonumber
&&{}= \int_0^T {\mathrm d}t\int_0^t {\mathrm d}s \, {\mathrm e}^{-\left( \frac
\gamma 2 + {\mathrm i}( \nu-\omega) \right)(t-s)}  \big\langle {\mathrm
e}^{\frac {\mathrm i} 2 \, \sigma_z \beta} \Psi(\xi,\lambda) \big|
\\ \nonumber
&&{}\times \widetilde U(t)^\dagger R(h)^\dagger \widetilde U(t) \widetilde
U(s)^\dagger R(h) \widetilde U(s)\, {\mathrm e}^{\frac {\mathrm i} 2 \,
\sigma_z \beta}\Psi(\xi,\lambda)\big\rangle,
\end{eqnarray}
where $R(h)$ is defined by Eq.\ (\ref{5.18c}) and
\[
\widetilde U(t) = {\mathrm e}^{\frac {\mathrm i} 2 \, \sigma_z( \beta +\omega
t)} U(t) {\mathrm e}^{-\frac {\mathrm i} 2 \, \sigma_z \beta}.
\]

By the quantum regression theorem, which holds for a dynamics like $\widetilde
U(t)$ \cite{Frig}, we have
\begin{eqnarray*}
\big\langle &&{\mathrm e}^{\frac {\mathrm i} 2 \, \sigma_z \beta}
\Psi(\xi,\lambda) \big|  \widetilde U(t)^\dagger R(h)^\dagger \widetilde U(t)
\widetilde U(s)^\dagger R(h) \widetilde U(s)\, {\mathrm e}^{\frac {\mathrm i} 2
\, \sigma_z \beta}
\\
&&{}\times\Psi(\xi,\lambda)\big\rangle = {\mathrm Tr} \left\{ R(h)^\dagger \,
{\mathrm e}^{{\mathcal L}_\lambda (t-s) } \left[ R(h) \, {\mathrm e}^{
{\mathcal L}_\lambda s} \left[ \rho_0\right] \right] \right\},
\end{eqnarray*}
where $\rho_0 = \exp \left( \frac {\mathrm i} 2 \sigma_z \beta\right)
|\xi\rangle \langle \xi |  \exp \left(- \frac {\mathrm i} 2 \sigma_z \beta
\right)$. By recalling that $\lim_{t\to +\infty} {\mathrm e}^{ {\mathcal
L}_\lambda t} [ \rho]= \rho_{\mathrm eq}^\lambda$ for any state $\rho$, we
obtain
\begin{eqnarray}\nonumber
\lim_{T\to +\infty} &&\frac{k}{ 2\pi T} \Big\langle \int_0^T {\mathrm d}
A^\dagger_h(t) \int_0^t {\mathrm d} A_h(s) \, {\mathrm e}^{ - \left( \frac
\gamma 2 +{\mathrm i} \nu \right) (t-s) } \Big\rangle_\lambda^T
\\ \nonumber
&&{}= \frac k {2\pi} \int_0^{+\infty } {\mathrm d}t\, {\mathrm e}^{ -\left(
\frac \gamma 2 +{\mathrm i}(\nu -\omega) \right) t}
\\ \label{aa}
&&{}\times {\mathrm Tr} \left\{ R(h)^\dagger \, {\mathrm e}^{{\mathcal
L}_\lambda t } [ R(h) \rho_{\mathrm eq}^\lambda] \right\}.
\end{eqnarray}
By inserting $R(h)=D(h)+r(h)$ into Eq.\ (\ref{aa}) and this equation into Eq.\
(\ref{5.8}), we obtain the decomposition (\ref{5.13})-(\ref{5.16}).

The positivity of $k/(4\pi)$ and $P_{\mathrm el}(\nu,h)$ is apparent from their
definitions, while to prove the positivity of $P_{\mathrm inel}(\nu,h)$
requires some transformations.

By repeating in the reverse order the steps from Eq.\ (\ref{bb}) to Eq.\
(\ref{aa}), we obtain from Eq.\ (\ref{5.16})
\begin{eqnarray*}
P_{\mathrm inel}(\nu,h) &=& \lim_{T\to +\infty} \frac 1 T \int_0^T {\mathrm d}t
\int_0^t {\mathrm d}s \, {\mathrm e}^{-\frac \gamma 2 |t-s|} \langle \phi(t) |
\phi(s) \rangle
\\ &+&
\lim_{T\to +\infty} \frac 1 T \int_0^T {\mathrm d}t \int_0^t {\mathrm d}s \,
{\mathrm e}^{-\frac \gamma 2 |t-s|} \langle \phi(s) | \phi(t) \rangle\,,
\end{eqnarray*}
where
\[
\phi(t) = \sqrt{ \frac k {2\pi}}\, {\mathrm e}^{{\mathrm i}(\nu - \omega)t}\,
\widetilde U(t)^\dagger D(h)\widetilde U(t) {\mathrm e}^{\frac {\mathrm i}{2}
\sigma_z \beta} \Psi(\xi,\lambda)\,.
\]
By exchanging the order of integration and the names of the variables $s$ and
$t$ in the second term, we get
\[
P_{\mathrm inel}(\nu,h) = \lim_{T\to +\infty} \frac 1 T \int_0^T {\mathrm d}t
\int_0^T {\mathrm d}s \, {\mathrm e}^{-\frac \gamma 2 |t-s|} \langle \phi(t) |
\phi(s) \rangle\,,
\]
which is positive because $\exp\left( -\frac \gamma 2 \left|t\right|\right)$ is
a positive-definite function, i.e.\ the Fourier transform of a positive
function. \EndProof

Notice that in the decomposition (\ref{5.13}) the term $k/(4\pi)$, independent
of $\nu$, is apparently a white noise contribution to the power; $P_{\mathrm
el}(\nu,h)$ is the elastic contribution, as one sees from Eq.\ (\ref{5.17})
which gives $P_{\mathrm el}(\nu,h)\propto \delta(\nu-\omega)$ for $\gamma
\downarrow 0$; finally, $P_{\mathrm inel}(\nu,h) $ is the inelastic
contribution (from Eq.\ (\ref{5.16}) one can see that no delta term develops
for $\gamma \downarrow 0$).

By Eqs.\ (\ref{3.10}), (\ref{3.27})-(\ref{3.29}), (\ref{5.17})-(\ref{5.18}), we
obtain
\begin{eqnarray}\nonumber
r(h)&=& \langle h| S^- \lambda \rangle +  \langle h| \Delta S \lambda \rangle
u(\infty)
\\
&&{}+ {\mathrm e}^{-{\mathrm i}\beta} \langle h| \alpha \rangle v(\infty)\,,
\label{5.12}
\end{eqnarray}
\begin{equation}\label{5.14}
P_{\mathrm inel} (\nu,h) = \frac k {2\pi} \, \bbox{c}^{h\,\dagger} \, \frac 1
{\bbox{G} + \frac \gamma 2 + {\mathrm i} (\nu-\omega)}\, \bbox{d}^{h} +
\text{c.c.}\,,
\end{equation}
\begin{equation}\label{5.15}
\bbox{c}^h = \left( \begin{array}{c} \langle h| \Delta S \lambda \rangle
\\ 0 \\ {\mathrm e}^{-{\mathrm i}\beta} \langle h| \alpha \rangle \end{array} \right),
\end{equation}
\begin{mathletters}\label{5.19}
\begin{eqnarray}\nonumber
d_1^h &=&  \bigl[ \langle h| \Delta S \lambda \rangle \big(1-u(\infty)\big)
\\ \label{5.19a}
&&{}- {\mathrm e}^{-{\mathrm i}\beta} \langle h| \alpha \rangle v(\infty)
\bigr] u(\infty)\,,
\\ \nonumber
d_2^h &=&  \bigl[ \langle h| \Delta S \lambda \rangle \big(1-u(\infty)\big)
\\
&&{}- {\mathrm e}^{-{\mathrm i}\beta} \langle h| \alpha \rangle v(\infty)
\bigr] v(\infty)\,, \label{5.19b}
\\ \nonumber
d_3^h &=& {\mathrm e}^{-{\mathrm i}\beta} \langle h| \alpha \rangle
\left(u(\infty) -|v(\infty)|^2  \right)
\\
&&{}-  \langle h| \Delta S \lambda \rangle u(\infty)\, \overline{v(\infty)}\,.
\label{5.19c}
\end{eqnarray}
\end{mathletters}

\subsection{Elastic and inelastic cross sections}\label{cross}
Let us consider now the case of the spherically symmetric atom, stimulated by a
well collimated laser beam, for which Eqs.\ (\ref{3.30})-(\ref{3.36}) hold. We
also assume that the detector spans a small solid angle, so that $h$ is given
by Eq.\ (\ref{5.4}) with $\Delta \Upsilon \downarrow \{(\theta,\phi)\}$, $|
\Delta \Upsilon | \simeq \sin \theta \, {\mathrm d }\theta \, {\mathrm d}
\phi$. Moreover, we assume that the transmitted wave does not reach the
detector, i.e.\ $\theta>0$ and so
\begin{equation}\label{5.22}
\langle h | \lambda\rangle =0\,.
\end{equation}
From Eqs.\ (\ref{5.17}), (\ref{5.12})-(\ref{5.19}) we obtain the elastic and
inelastic contributions to the power (per unit of solid angle)
\begin{mathletters}\label{5.23}
\begin{eqnarray}\label{5.23a}
\frac 1 {|\Delta \Upsilon|} \, P_{\mathrm el}(\nu,h) &\simeq&  P_{\mathrm
el}(\nu;\theta,\phi)\,,
\\ \label{5.23b}
\frac 1 {|\Delta \Upsilon|} \, P_{\mathrm inel}(\nu,h) &\simeq&  P_{\mathrm
inel}(\nu;\theta,\phi)\,,
\end{eqnarray}
\end{mathletters}
\noindent where
\begin{mathletters}\label{5.23'}
\begin{equation}\label{5.23'a}
P_{\mathrm el}(\nu;\theta,\phi) = k \eta^2 \|\alpha\|^2 |a(\theta)|^2\, \frac
{\gamma/(2\pi)} {(\nu - \omega)^2 + \gamma^2/4}\,,
\end{equation}
\begin{eqnarray}\nonumber
P_{\mathrm inel}(\nu;\theta,\phi) &=& \frac{ k\eta^2 \|\alpha\|^2} {2\pi}
\bbox{c}(\theta)^\dagger \, \frac 1 {\bbox{G} + \frac \gamma 2 + {\mathrm i}
(\nu-\omega)}\, \bbox{d}(\theta)
\\ &&{}+ \text{c.c.}\,,\label{5.23'b}
\end{eqnarray}
\end{mathletters}
\begin{eqnarray}\nonumber
a(\theta)&=& g_-(\theta) + \Delta g(\theta)\, \frac { \eta^2 \kappa^2}
{z^2+\zeta^2}
\\ \label{5.24}
&&{}- {\mathrm e}^{2{\mathrm i}\delta_0^-} \, \frac {\kappa^2 + {\mathrm i} y }
{\sqrt{4\pi} \left(z^2+\zeta^2\right)} \, ,
\end{eqnarray}
\begin{equation}\label{5.25}
\bbox{c}(\theta) = \left( \begin{array}{c} \eta \Delta g(\theta)
\\ 0 \\
-{\mathrm e}^{2{\mathrm i}\delta_0^-} /\sqrt{4\pi} \end{array} \right),
\end{equation}
\begin{mathletters}\label{5.26}
\begin{eqnarray}\label{5.26a}
d_1(\theta)&=& \frac {\eta\kappa^2} {z^2+\zeta^2}\, m(\theta)\,,
\\ \label{5.26b}
d_2(\theta)&=& \frac {m(\theta)} {z^2+\zeta^2} \left( \kappa^2 + {\mathrm i}
y\right) ,
\\ \nonumber
d_3(\theta) &=& -\frac {\eta^2 } { \left( z^2 +\zeta^2\right)^2} \biggl\{
\frac{{\mathrm e}^{2{\mathrm i}\delta_0^-}}{\sqrt{4\pi}} \Bigl[ \|\Delta g\|^2
\left(y^2+ \kappa^4\right)
 \\ \nonumber
&&{}+ \kappa^2  y\sin 2 s + 2\kappa^4 \cos^2 s  \Bigr]
\\ \label{5.26c}
&&{}+ \Delta g(\theta) \kappa^2 \left( \kappa^2 - {\mathrm i}y \right)
\biggr\},
\end{eqnarray}
\end{mathletters}
\begin{eqnarray}\nonumber
m(\theta)&=&  \Delta g(\theta) \left( 1 - \frac {\eta^2 \kappa^2} {z^2
+\zeta^2} \right)
\\ \label{5.27}
&&{}+{\mathrm e}^{2 {\mathrm i} \delta_0^-} \,  \frac { \kappa^2+{\mathrm i}y}
{\sqrt{4\pi} \left( z^2 +\zeta^2\right)}\,.
\end{eqnarray}

For the elastic and inelastic cross sections we shall have $\sigma_{\mathrm
el}(\nu;\theta,\phi) \propto P_{\mathrm el}(\nu;\theta,\phi)$, $\sigma_{\mathrm
inel}(\nu;\theta,\phi) \propto P_{\mathrm inel}(\nu;\theta,\phi)$. To find the
constant of proportionality, let us observe that, from Eqs.\ (\ref{5.21}),
(\ref{5.23}), (\ref{4.1}), (\ref{4.3}), (\ref{4.5}), we get
\begin{eqnarray}\nonumber
\int_{-\infty}^{+\infty}&& \left[P_{\mathrm el}(\nu;\theta,\phi) + P_{\mathrm
inel} (\nu;\theta,\phi)\right] {\mathrm d} \nu
\\ \label{5.29}
{}=&& \lim_{t\to +\infty} \frac k t \, \langle n(\theta, \phi;t\rangle =
\frac{\eta^2 k \|\alpha\|^2 \omega^2} {6\pi c^2}\, \sigma(\theta,\phi)\,.
\end{eqnarray}
Therefore, taking into account Eqs.\ (\ref{5.23'}), we obtain the expressions
for the cross sections
\begin{equation}\label{5.30}
\sigma_{\mathrm el}(\nu;\theta,\phi)= \frac{3 c^2}{ \omega^2} \left|
a(\theta)\right|^2\,\frac {\gamma} {(\nu - \omega)^2 + \gamma^2/4}\,,
\end{equation}
\begin{equation}\label{5.31}
\sigma_{\mathrm inel}(\nu;\theta,\phi)=  \frac {3 c^2} {\omega^2} \, \bbox{
c}(\theta)^\dagger \, \frac {1} {\bbox{G} + \frac \gamma 2 + {\mathrm i}
(\nu-\omega)}\, \bbox{d}(\theta) + \text{c.c.}\,,
\end{equation}
and the relation
\begin{equation}\label{5.32}
\int_{-\infty}^{+\infty} \left[ \sigma_{\mathrm el}(\nu; \theta,\phi) +
\sigma_{\mathrm inel}(\nu;\theta,\phi)\right] {\mathrm d}\nu =
\sigma(\theta,\phi)\,,
\end{equation}
where $\sigma(\theta,\phi)$ is given by Eq.\ (\ref{4.9}).

Finally, let us introduce the integral cross sections
\begin{mathletters}\label{5.33}
\begin{eqnarray}\label{5.33a}
\sigma_{\mathrm el}(\nu) &=& \int_0^\pi {\mathrm d}\theta \, \sin \theta
\int_0^{2\pi} {\mathrm d}\phi \, \sigma_{\mathrm el}(\nu;\theta,\phi)\,,
\\ \label{5.33a'}
\sigma_{\mathrm el} &=& \int_{-\infty}^{+\infty} \sigma_{\mathrm el}(\nu)
\,{\mathrm d}\nu\,,
\\ \label{5.33b}
\sigma_{\mathrm inel}(\nu) &=& \int_0^\pi {\mathrm d}\theta \, \sin \theta
\int_0^{2\pi} {\mathrm d}\phi \, \sigma_{\mathrm inel}(\nu;\theta,\phi)\,,
\\ \label{5.33c}
\sigma_{\mathrm inel} &=& \int_{-\infty}^{+\infty} \sigma_{\mathrm inel}(\nu)
\,{\mathrm d}\nu\,;
\end{eqnarray}
\end{mathletters}
\noindent the relation (\ref{5.32}) becomes
\begin{equation}\label{tot}
\sigma_{\mathrm el}+\sigma_{\mathrm inel} =\sigma_{{}_{\mathrm TOT}}\,,
\end{equation}
where $\sigma_{{}_{\mathrm TOT}}$ is given by Eqs.\ (\ref{4.12})-(\ref{4.14}).

\section{Cross sections and fluorescence spectrum}\label{Dis}
In this section we want to discuss the behavior of the integral cross sections
and of the fluorescence spectrum.

From Eqs.\ (\ref{5.24}), (\ref{5.30}), (\ref{5.33a}), (\ref{5.33a'}) we obtain
\begin{equation}\label{6.1-}
\sigma_{\mathrm el}(\nu)=  \sigma_{\mathrm el}\, \frac{\gamma/(2\pi)} {(\nu
-\omega)^2 + \gamma^2/4}\,,
\end{equation}
\begin{eqnarray}\nonumber
\frac{\omega^2}{6\pi c^2}&&\, \sigma_{\mathrm el} = \frac 1
{\left(z^2+\zeta^2\right)^2} \left\| P_\bot \left[ \left( z^2 +B \right) g_- +
\eta^2 \kappa^2 g_+ \right] \right\|^2
\\ \label{6.1}
&&{}+ \left| {\mathrm e}^{-{\mathrm i}\delta_0^-} \sin \delta_0^- + \frac
{\eta^2 \kappa^2 {\mathrm e}^{{\mathrm i}s} \sin s - y + {\mathrm i}\kappa^2} {
z^2 + \zeta^2} \right|^2\,,
\end{eqnarray}
while from Eqs.\ (\ref{5.25})-(\ref{5.27}), (\ref{5.31}), (\ref{5.33b}),
(\ref{5.33c}) we obtain
\begin{equation}\label{6.2}
\frac{\omega^2}{6\pi c^2}\, \sigma_{\mathrm inel} = \frac {\eta^2 \left( 1 +
\kappa^2 \right) E(y)} { \left(z^2 +\zeta^2\right)^2}\,,
\end{equation}
\begin{equation}\label{6.3}
E(y)= \left(y \sin s + \kappa^2 \cos s \right)^2 + \left\| P_\bot \Delta g
\right\|^2 \left(y^2 + \kappa^4\right);
\end{equation}
one can check that the relation (\ref{tot}) holds true.

Let us recall that the various quantities appearing in the previous formulas
are given by Eqs.\ (\ref{3.7i}), (\ref{3.34})-(\ref{5.28}), (\ref{3.36m}),
(\ref{3.36c}), (\ref{4.13}) and that $\|\alpha\|^2$ is the natural line width,
$\omega_0$ is the atomic resonance frequency, $\Omega =\eta \|\alpha\|^2$ is
the bare Rabi frequency, $\eta^2 \varepsilon$ is the intensity dependent shift,
$\delta_0^\pm$, $\left\| P_\bot g_\pm \right\|^2$, $\left\| P_\bot \Delta g
\right\|^2$ are parameters linked to the $S_\pm$ scattering matrices,
satisfying
\[
\Big| \left\| P_\bot g_+ \right\| - \left\| P_\bot g_- \right\| \Big| \leq
\left\| P_\bot \Delta g \right\| \leq  \left\| P_\bot g_+ \right\| + \left\|
P_\bot g_- \right\| .
\]
We introduce also a ``reduced" detuning $\widetilde z$
\begin{equation}\label{6.red}
\widetilde z = \frac{ \omega - \omega_0} {\|\alpha\|^2}\, ;
\end{equation}
we have also $z= 2 \widetilde z - 2 \eta^2 \varepsilon / \|\alpha\|^2$, $s=
\delta_0^+ - \delta_0^-$, $y = z- \frac{\eta^2} 2 \sin 2s$.

As an example, in Fig.\ \ref{fig1} we plot $\frac{\omega^2}{6\pi c^2}\,
\sigma_{{}_{\rm TOT}}$, $\frac{\omega^2}{6\pi c^2}\, \sigma_{\rm inel}$,
$\frac{\omega^2}{6\pi c^2}\, \sigma_{\rm el}$ as functions of the detuning
$\widetilde z$ in the four cases $\eta^2 = 10,\, 18,\, 28,\, 40$; the other
parameters are $\delta_0^+=-0.03$, $\delta_0^-=0.13$, $\left\| P_\bot g_\pm
\right\|^2 = 0.005$, $\left\| P_\bot \Delta g \right\|^2= 0.02$, $\varepsilon/
\|\alpha\|^2=-0.001$. Let us note the strong asymmetry in $\widetilde z$ of the
cross sections and the fact that $\displaystyle \lim_{\widetilde z \to \pm
\infty} \frac{\omega^2}{6\pi c^2}\, \sigma_{{}_{\rm TOT}} = \lim_{\widetilde z
\to \pm \infty} \frac{\omega^2}{6\pi c^2}\, \sigma_{\rm el} = \left\| P_\bot
g_- \right\|^2 + \sin^2 \delta_0^-$, which is about 0.0218 with our parameters.

Let us recall that the usual model with only the absorption/emission process
corresponds to $\delta_0^\pm=0$, $\left\| P_\bot g_\pm \right\|^2=\left\|
P_\bot \Delta g \right\|^2= 0$, $\varepsilon=0$, $z=2\widetilde z$; in this
case, from Eqs.\ (\ref{4.11}), (\ref{6.1}), (\ref{6.2}), we have easily
\begin{mathletters}\label{5.48}
\begin{eqnarray}\label{5.48a}
\frac{\omega^2}{6\pi c^2}\, \sigma_{{}_{\mathrm TOT}}&=& \frac{1}{4 \widetilde
z^2+1+2\eta^2}\,,
\\ \label{5.48b}
\frac{\omega^2}{6\pi c^2}\,\sigma_{\mathrm el}&=&  \frac{4 \widetilde z^2
+1}{\left(4 \widetilde z^2 +1+2\eta^2\right)^2}\,,
\\ \label{5.48c}
\frac{\omega^2}{6\pi c^2}\,\sigma_{\mathrm inel} &=& \frac{2\eta^2}{\left(4
\widetilde z^2+1+2\eta^2\right)^2}\,.
\end{eqnarray}
\end{mathletters}
\noindent Now the cross sections are symmetric in $\widetilde z$ and
$\displaystyle \lim_{\widetilde z \to \pm \infty} \frac{\omega^2}{6\pi c^2}\,
\sigma_{{}_{\rm TOT}} = \lim_{\widetilde z \to \pm \infty} \frac{\omega^2}{6\pi
c^2}\, \sigma_{\rm el} =0$.


Then, we introduce the normalized inelastic spectrum
\begin{equation}\label{6.4}
\Sigma_{\mathrm inel}(x) = \frac{\omega^2 \|\alpha\|^2} {6\pi c^2} \,
\sigma_{\mathrm inel}(\nu)
\end{equation}
and the total one
\begin{equation} \label{6.21}
\Sigma_{{}_{\mathrm TOT}}(x)=  \frac {\omega^2 \sigma_{\mathrm el}} {6\pi c^2}
\, \frac {\widetilde \gamma /(2\pi)} {x^2 + (\widetilde \gamma/2)^2} +
\Sigma_{\mathrm inel}(x)\,,
\end{equation}
where we have introduced the ``reduced" frequency $x$ and the ``reduced"
instrumental width $\widetilde \gamma$
\begin{equation}\label{6.5}
x= \frac{ \nu-\omega} {\|\alpha\|^2}\,,  \qquad \widetilde \gamma = \frac
\gamma {\|\alpha\|^2}\,;
\end{equation}
the normalization we have chosen is
\begin{mathletters}\label{6.15}
\begin{eqnarray}\label{6.15a}
\int_{-\infty}^{+\infty} \Sigma_{{}_{\mathrm TOT}}(x)\, {\mathrm d }x &=&
\frac{\omega^2}{6\pi c^2}\, \sigma_{{}_{\rm TOT}}\,,
\\ \label{6.15b}
\int_{-\infty}^{+\infty} \Sigma_{\mathrm inel}(x)\, {\mathrm d }x &=&
\frac{\omega^2}{6\pi c^2}\, \sigma_{\mathrm inel}\,.
\end{eqnarray}
\end{mathletters}
\noindent The explicit expression of $\Sigma_{\mathrm inel}(x)$ is given by the
following proposition.

\begin{proposition}\label{prop6}
The inelastic spectrum is given by
\begin{eqnarray}\nonumber
\Sigma_{\mathrm inel}(x) &=& \frac {\eta^2} {\pi \left(z^2+\zeta^2\right)^2}
\biggr( \bbox{c}^{\prime \dagger} \, \frac 1 {\widetilde{\bbox{G}} + 2{\mathrm
i} x}\, \bbox{d}^\prime
\\ \label{6.9}
&&{}+ \left\|P_\bot \Delta g \right\|^2 \bbox{c}^{\prime\prime \dagger} \,
\frac 1 {\widetilde{\bbox{G}} + 2{\mathrm i} x}\, \bbox{d}^{\prime\prime} +
\text{\rm c.c.} \biggr),
\end{eqnarray}
where
\begin{equation}\label{6.6}
\bbox{c}^\prime = \left( \begin{array}{c} {\mathrm i} {\mathrm e}^{{\mathrm i}
s} \sin s
\\ 0 \\
1 \end{array} \right), \qquad \bbox{c}^{\prime\prime} = \left(
\begin{array}{c} 1\\ 0 \\ 0 \end{array} \right),
\end{equation}
\begin{mathletters}\label{6.7}
\begin{eqnarray}\label{6.7a}
d_1^\prime&=& \kappa^2 m^\prime\,,
\\ \label{6.7b}
d_2^\prime&=& \left( \kappa^2 + {\mathrm i} y\right) m^\prime\,,
\\ \nonumber
d_3^\prime &=& \|\Delta g\|^2 \left(y^2+ \kappa^4\right) + \kappa^2 y\sin 2 s
\\ \label{6.7c}
&&{}+ 2\kappa^4 \cos^2 s + {\mathrm i} \kappa^2 \left( \kappa^2 - {\mathrm i}y
\right) {\mathrm e}^{{\mathrm i}s}\sin s   \,,
\\ \label{6.7g}
m^\prime &=& \kappa^2 + {\mathrm i} y + {\mathrm i}\left( z^2 + \zeta^2 -
\eta^2\kappa^2 \right) {\mathrm e}^{{\mathrm i}s} \sin s \,,
\\ \label{6.7d}
d_1^{\prime\prime}&=& \kappa^2 \left(z^2+ \zeta^2 -\eta^2 \kappa^2 \right),
\\ \label{6.7e}
d_2^{\prime\prime} &=& \left( \kappa^2 + {\mathrm i} y\right) \left(z^2+
\zeta^2 -\eta^2 \kappa^2 \right),
\\ \label{6.7f}
d_3^{\prime\prime}&=& \kappa^2 \left( \kappa^2 - {\mathrm i}y \right) ,
\end{eqnarray}
\end{mathletters}
\begin{equation}\label{6.8}
\widetilde{\bbox{G}}= \left( \begin{array}{ccc} 2 + \widetilde \gamma & - 1 &
\eta^2
 \\ & & \\
2\eta^2 {\mathrm e}^{{\mathrm i}s} \cos s  &
 b^\prime + \widetilde \gamma & 0 \\ & & \\
-2{\mathrm e}^{-{\mathrm i}s} \cos s & 0
 &  \overline{ b^\prime} + \widetilde \gamma \end{array} \right);
\end{equation}
$b^\prime$ is given by Eq.\ (\ref{set5}).
\end{proposition}

\Proof  From Eqs.\ (\ref{3.36a}), (\ref{5.25}), (\ref{5.26}), (\ref{5.27}),
(\ref{5.31}), (\ref{6.2}), (\ref{6.4}), (\ref{6.5}),  we have
\begin{eqnarray*}
\Sigma_{\mathrm inel}(x) &=& \frac 1 {\pi \left(z^2+\zeta^2\right)^2}
\biggr(\widetilde {\bbox{c}}^{\prime \dagger} \, \frac 1 {\bbox{G}^\prime +
\widetilde \gamma+ 2{\mathrm i} x}\, \widetilde{\bbox{d}}^\prime
\\
&&{}+ \eta \left\|P_\bot \Delta g \right\|^2 \bbox{c}^{\prime\prime \dagger} \,
\frac 1 {\bbox{G}^\prime + \widetilde \gamma+ 2{\mathrm i} x}\, \widetilde{
\bbox{d}}^{\prime\prime} + \text{c.c.} \biggr),
\end{eqnarray*}
where, by using the definitions above,
\[
\widetilde{\bbox{c}}^\prime = \left( \begin{array}{c} {\mathrm i} \eta{\mathrm
e}^{{\mathrm i} s} \sin s
\\ 0 \\
-1 \end{array} \right),
\]
\[
\widetilde{\bbox{d}}^\prime = \left( \begin{array}{c} \eta d_1^\prime
\\ d_2^\prime \\
-\eta^2 d_3^\prime \end{array} \right), \qquad \widetilde {\bbox{
d}}^{\prime\prime} = \left( \begin{array}{c} \eta d_1^{\prime\prime}
\\ d_2^{\prime\prime} \\
-\eta^2 d_3^{\prime\prime} \end{array} \right).
\]

Then, by using the transformation
\[
\bbox{G}^\prime+ \widetilde \gamma= \bbox{M}\widetilde{ \bbox{ G}} \bbox{
M}^{-1}\,,
\]
where
\[
\bbox{ M}= \left( \begin{array}{ccc} \eta & 0 & 0
\\ 0 & 1 &0 \\ 0 & 0 &
-\eta^2 \end{array} \right),
\]
we get the Eq.\ (\ref{6.9}). \EndProof

By direct computations we can get the inverse of the matrix $\widetilde {\bbox{
G}}+2{\mathrm i}x$. We can write
\begin{equation}\label{6.16}
\frac{1}{\widetilde{ \bbox{ G}} + 2{\mathrm i} x} = \frac{1}{\det
\left(\widetilde{ \bbox{ G}} + 2{\mathrm i} x\right)}\, \bbox{ D}(x)\,,
\end{equation}
where
\end{multicols}
\vspace{-0.4cm}
\noindent\rule{0.49\textwidth}{0.4pt}\rule{0.4pt}{\baselineskip}
\widetext
\begin{eqnarray}\nonumber
\det \left( \widetilde{ \bbox{ G}} + 2{\mathrm i} x\right)&=& \left(2+
\widetilde \gamma+2{\mathrm i}x \right) \left[ \left( \kappa^2 + \widetilde
\gamma+ 2{\mathrm i}x \right)^2 +  \left( z + \frac {\eta^2} 2 \, \sin 2s
\right)^2 \right]
\\
&&{}+ 4 \eta^2 \cos s \left[ \left(\kappa^2 + \widetilde \gamma+ 2{\mathrm i}x
\right)\cos s - \left( z + \frac {\eta^2} 2 \, \sin 2s \right)\sin s \right],
\label{6.17}
\end{eqnarray}
\begin{mathletters}\label{6.18}
\begin{eqnarray}\label{6.18a}
D_{11}(x)&=&  \left( \kappa^2 + \widetilde \gamma+ 2{\mathrm i}x \right)^2 +
\left( z + \frac {\eta^2} 2 \, \sin 2s \right)^2,
\\ \label{6.18b}
D_{12}(x)&=& \kappa^2   + \widetilde \gamma+ {\mathrm i} \left(2x+ z + \frac
{\eta^2} 2 \, \sin 2s \right),
\\ \label{6.18c}
D_{13}(x)&=& -\eta^2 \left[\kappa^2 + \widetilde \gamma+  {\mathrm i} \left(2x-
z - \frac {\eta^2} 2 \, \sin 2s \right)\right],
\\ \label{6.18g}
D_{31}(x)&=&  2 {\mathrm e}^{-{\mathrm i}s}\, \cos s\left[\kappa^2 + \widetilde
\gamma+ {\mathrm i} \left(2x- z - \frac {\eta^2} 2 \, \sin 2s \right)\right]\,,
\\ \label{6.18h}
D_{32}(x)&=&  2 {\mathrm e}^{-{\mathrm i}s}\, \cos s\,,
\\ \label{6.18i}
D_{33}(x)&=&  \left(2+ \widetilde \gamma+2 {\mathrm i} x\right)\left[\kappa^2 +
\widetilde \gamma+ {\mathrm i} \left(2x- z - \frac {\eta^2} 2 \, \sin 2s
\right)\right]+2\eta^2 {\mathrm e}^{{\mathrm i}s}\, \cos s\,.
\end{eqnarray}
\end{mathletters}
\begin{multicols}{2}
\narrowtext
\noindent The matrix elements $D_{2j}(x)$ are not needed in formula
(\ref{6.9}).

One can check that the inelastic spectrum is asymmetric, but it is invariant
under the transformation: $x\to - x$, $s\to - s$, $z\to - z$.

The formula for the inelastic spectrum given in Proposition \ref{prop6} becomes
significantly simpler in the usual case ($g_\pm =0$) and when the intensity of
the stimulating laser is low.

\subsubsection*{The case $g_\pm =0$}
Let us consider the usual model, when the direct scattering terms are
negligible, i.e.\ $g_\pm=0$, which gives also $s=0$, $\zeta^2=1+2\eta^2$,
$\kappa^2=1$, $y=z=2\widetilde z$; in this case the integral cross sections are
given by Eqs.\ \ref{5.48} and, with some computations, the inelastic spectrum
is obtained from Proposition \ref{prop6}:
\begin{mathletters}\label{5.49}
\begin{equation} \label{5.49a}
\Sigma_{\mathrm inel}(x) = \frac {4\eta^2 p(x)} {\pi q(x)
\left(z^2+1+2\eta^2\right)^2}\,,
\end{equation}
\begin{eqnarray}\nonumber
p(x)&=& \left( 2 + \widetilde \gamma  \right) \left[ \left( 1  + \widetilde
\gamma  \right)^2 +2 \eta^2 + z^2 \right]
\\ \label{5.49b}
{}&\times& \left[ \left( 2 + \widetilde \gamma \right)^2 + 2 \eta^2 +4x^2
\right]
\\  \nonumber {}&+&
2\widetilde \gamma  \left[ 2\left(2 x^2 - \eta^2\right)^2 + \left( 2 +
\widetilde \gamma  \right)^2\left(2 x^2 + \eta^2 \right) \right],
\end{eqnarray}
\begin{eqnarray}\nonumber
q(x)&=& \biggl\{\left( 2 + \widetilde \gamma  \right) \left[ \left( 1 +
\widetilde \gamma  \right)^2 + z^2 \right]
\\ \label{5.49c}
{}&+& 4 \left( 1 + \widetilde \gamma \right)\eta^2 - 4\left( 4 +  3  \widetilde
\gamma  \right) x^2 \biggr\}^2
\\ \nonumber
{}&+& 4x^2 \left(3 \widetilde \gamma^2 +8 \widetilde \gamma +5 + z^2 + 4 \eta^2
- 4 x^2\right)^2  .
\end{eqnarray}
\end{mathletters}
\noindent Now the inelastic spectrum is invariant either under the
transformation $x\to - x$ either under the transformation $z\to - z$.

If we put also $\widetilde \gamma=0$, which means that the instrumental width
is negligible, then one can check that the fluorescence spectrum
$\Sigma_{{}_{\mathrm TOT}}(x)$, given by Eqs.\ (\ref{6.21}), (\ref{5.48}),
(\ref{5.49}), coincides exactly (apart from the different normalization) with
the spectrum computed by Mollow $\big(\!$~\cite{Moll}, Eq.\ (4.15)$\big)$. Eq.\
(\ref{5.49}) is simply the convolution of the inelastic part of the Mollow
spectrum with a Lorentzian of width $\widetilde \gamma$.

If also $z=0$ (no detuning), the eigenvalues of $\widetilde{ \bbox{ G}}$ can be
computed and, by using them, the denominator in Eq.\ (\ref{5.49}) can be
factorized. In the case $\eta^2\leq 1/16$, $\widetilde{ \bbox{ G}}$ has real
eigenvalues and $\Sigma_{\mathrm inel}(x)$ has a single peak in $x=0$, while
for $\eta^2>1/16$ two complex eigenvalues appear; therefore, $\Sigma_{\mathrm
inel}(x)$ has a three-peaked structure for $\eta^2$ sufficiently larger than
$1/16$. For $\eta$ very large Eq.\ (\ref{5.49}) gives three peaks in $\nu\simeq
\omega- \Omega$, $\nu=\omega=\omega_0$, $\nu\simeq \omega+ \Omega$ with height
ratio $1 : \frac{3+2\widetilde \gamma} {1+ \widetilde \gamma} :1$ and widths
$\frac 3 2 \, \|\alpha\|^2 +  \gamma $, $\|\alpha\|^2+ \gamma $, $\frac 3 2 \,
\|\alpha\|^2+ \gamma $ (see Ref.\ \cite{Moll} or Ref.\ \cite{CT92} pp.\ 387,
423-426, 437-441 for the case $\gamma =0$).

\subsubsection*{Low intensity laser}
From Eq.\ (\ref{6.2}) we see that the inelastic cross section vanishes in the
limit of vanishing intensity of the laser; however, the first correction,
proportional to $\eta^2$, presents some interesting aspects. We have
immediately
\begin{equation}\label{5.37}
\frac {\omega^2}{6\pi c^2}\, \sigma_{\mathrm inel} \simeq \frac {2\eta^2
E_0\left( \widetilde z \right)} {\left(4 \widetilde z^2 + 1\right)^2}\,,
\end{equation}
\begin{eqnarray}\nonumber
E_0\left( \widetilde z \right) = E(y)\big|_{\eta=0}&=& \left(2 \widetilde z\sin
s + \cos s\right)^2
\\ &+&
\left\|P_\bot \Delta g\right\|^2 \left(4 \widetilde z^2+1\right).\label{6.22}
\end{eqnarray}
The computation of the spectrum is straightforward, but long; the final result
is: for small $\eta$ we have
\end{multicols}
\vspace{-0.4cm}
\noindent\rule{0.49\textwidth}{0.4pt}\rule{0.4pt}{\baselineskip}
\widetext
\begin{eqnarray}\nonumber
\Sigma_{\mathrm inel}(x) &\simeq& \frac{\eta^2} {2\pi} \left[
\frac{\left\|P_\bot \Delta g\right\|^2\left(1+ \widetilde \gamma\right)}
{\widetilde z^2 +1/4} +\frac {\widetilde \gamma \left( 2 \widetilde z \sin s +
\cos s \right)^2} {4 \left( \widetilde z^2+1/4 \right)^2} \right] \left[ \frac
1 {4\left(x +\widetilde z \right)^2 + \left(1+\widetilde \gamma \right)^2} +
\frac 1 {4\left(x -\widetilde z \right)^2 + \left(1+\widetilde \gamma
\right)^2} \right]
\\  \label{5.35}
&+&\frac{2\eta^2 \left(2 \widetilde z \sin s +  \cos s \right)^2 \left[
\widetilde z^2 + \left(1+ \widetilde \gamma \right)^2/4 \right]} {\pi \left(
\widetilde z^2 +1/4\right)^2 \left[4\left(x +\widetilde z\right)^2 + \left(1+
\widetilde \gamma \right)^2 \right] \left[4\left(x -\widetilde z\right)^2 +
\left(1+ \widetilde \gamma \right)^2 \right] } \,.
\end{eqnarray}
\noindent{}\hfill{
\rule[-\baselineskip]{0.4pt}{\baselineskip}\rule{0.49\textwidth}{0.4pt}}
\vspace{-0.4cm}
\begin{multicols}{2}
\narrowtext
In this case the inelastic spectrum is invariant either under the
transformation $x\to - x$ either under the transformation $s\to - s$ and
$\widetilde z\to - \widetilde z$.

The usual case $(g_\pm =0)$ was already discussed by Mollow $\big($\cite{Moll},
Eq.\ (4.30)$\big)$ for $\widetilde \gamma=0$ and can be obtained from Eqs.\
(\ref{5.49}) by letting $\eta^2$ vanish or from Eq.\ (\ref{5.35}) by taking
$s=0$ and $\left\|P_\bot \Delta g\right\|=0$. In the Mollow case, for $|\Delta
\omega|$ sufficiently large, the inelastic spectrum presents two peaks (see
also Ref.\ \cite{CT92}, pp.\ 106-108, 386). The structure given by Eq.\
(\ref{5.35}) is similar also for $s\neq 0$, $\left\|P_\bot \Delta g\right\|
\neq 0$: again two symmetric peaks appear for $|\Delta \omega|$ sufficiently
large.

\subsubsection*{Numerical computations}

\begin{figure}
\begin{center}
{\null \hskip 0.0cm\epsfysize10truecm \epsfxsize16truecm \epsfbox{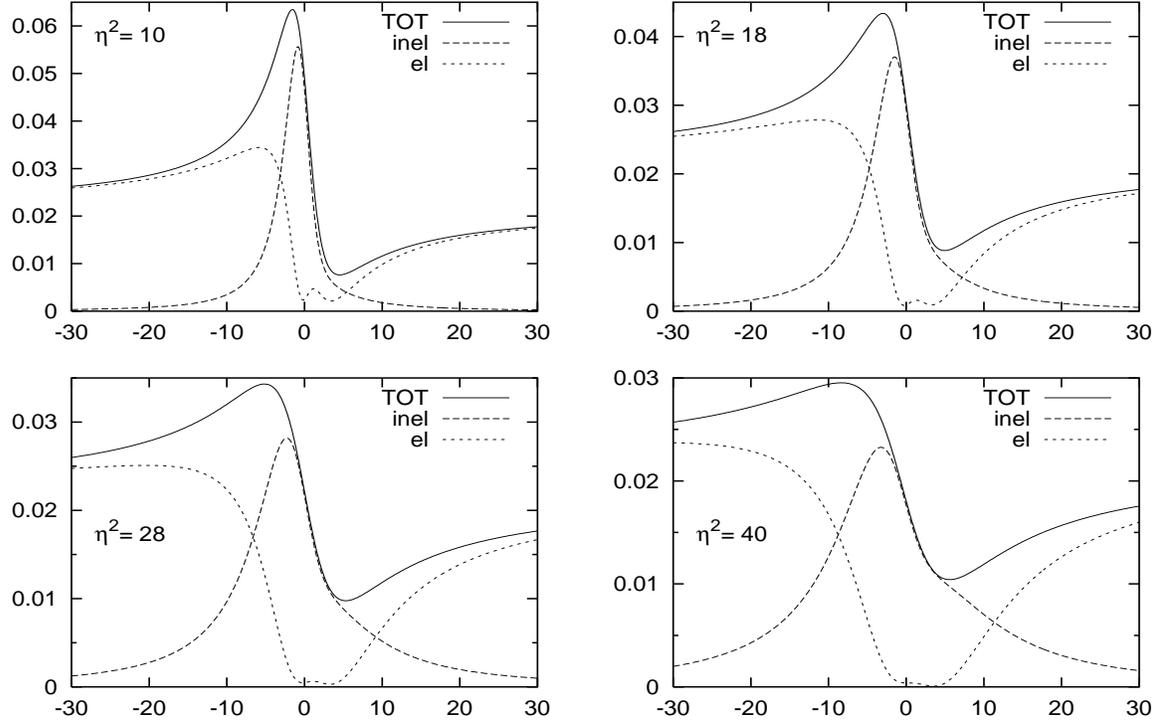}}
\caption{$\frac{\omega^2}{6\pi c^2}\times{}\,$the integral cross sections as
functions of the detuning $\widetilde z$ for $\delta_0^+=-0.03$,
$\delta_0^-=0.13$, $\left\| P_\bot g_\pm \right\|^2 = 0.005$, $\left\| P_\bot
\Delta g \right\|^2= 0.02$, $\varepsilon/ \|\alpha\|^2=-0.001$, and $\eta^2 =
10,\, 18,\, 28, \, 40$.} \label{fig1}
\end{center}
\end{figure}

\begin{figure}
\begin{center}
{\null \hskip 0.0cm\epsfysize10truecm \epsfxsize16truecm \epsfbox{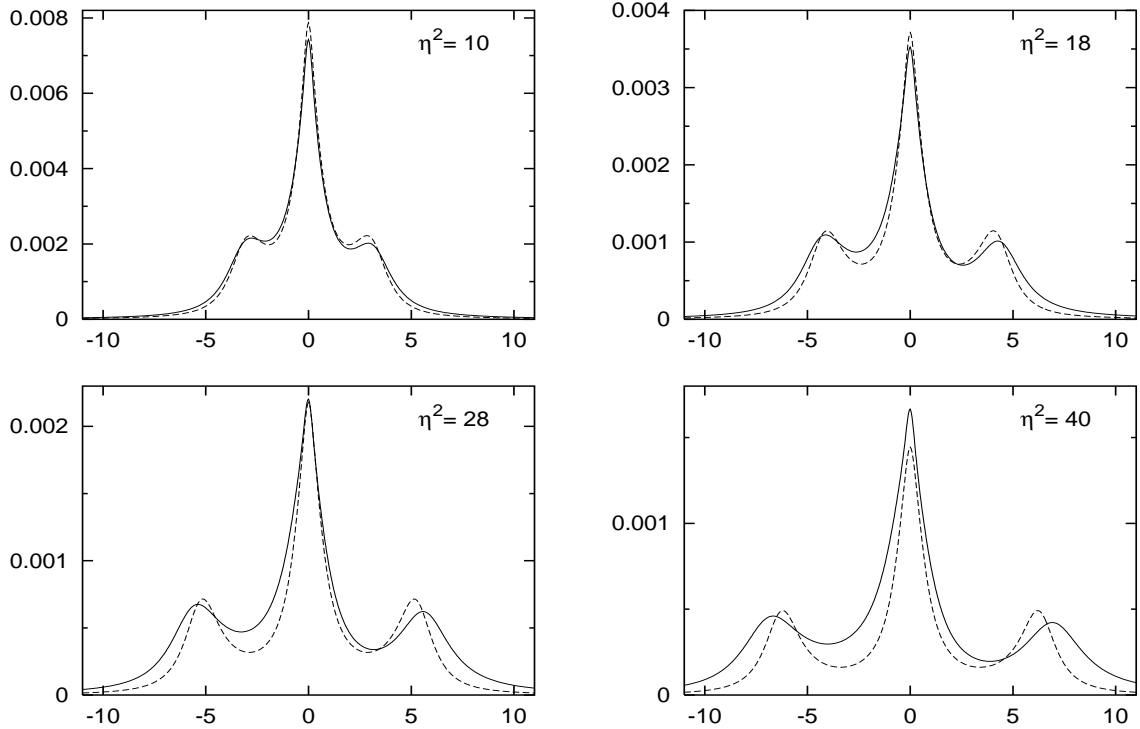}}
\caption{Total spectrum as a function of the frequency $x$ for $\widetilde
z=0$, $\widetilde \gamma=0.6$ and $\eta^2= 10,\, 18,\, 28, \, 40$; solid line:
$\delta_0^+=-0.03$, $\delta_0^-=0.13$, $\left\| P_\bot g_\pm \right\|^2 =
0.005$, $\left\| P_\bot \Delta g \right\|^2= 0.02$, $\varepsilon/
\|\alpha\|^2=-0.001$; dashed line: $\delta_0^\pm=0$, $\left\| P_\bot g_\pm
\right\|^2 = \left\| P_\bot \Delta g \right\|^2= 0$, $\varepsilon/
\|\alpha\|^2=0$.} \label{fig2}
\end{center}
\end{figure}

In the general case the total spectrum is given by Eqs.\ (\ref{6.21}),
(\ref{6.9})--(\ref{6.18}); the analytic expression is involved, but plots can
be easily obtained by numerical computations. According to the values of the
various parameters, a well resolved triplet structure can appear, but also
single-maximum structures can be shown. With the choice of parameters of Fig.\
\ref{fig1} and with an instrumental width $\widetilde \gamma = 0.6$, the on
resonance spectrum for $\eta^2=10,\, 18,\, 28,\, 40$ is given in Fig.\
\ref{fig2} (solid lines); the dashed lines give the Mollow spectrum for the
same values of $\eta^2$ and $\widetilde \gamma$. The parameters in Fig.\
\ref{fig2} have been chosen in such a way that a triplet structure appears, not
too different from the usual one, but with a well visible asymmetry in the
frequency $x$. Experiments \cite{Stroud,Walther1,Eze2,Eze1,Walther2} confirm
essentially the triplet structure; some asymmetry has been found, whose origin
has been attributed to various causes. In this connection it has also been
observed that calculations for multilevel atoms indicate some asymmetry
\cite{Eze2}. Indeed, the introduction in our model of the interaction term
containing the gauge process simulates the presence of other levels and the
virtual transitions to them.

\begin{figure}
\begin{center}
{\null \hskip 0.0cm\epsfysize10truecm \epsfxsize16truecm \epsfbox{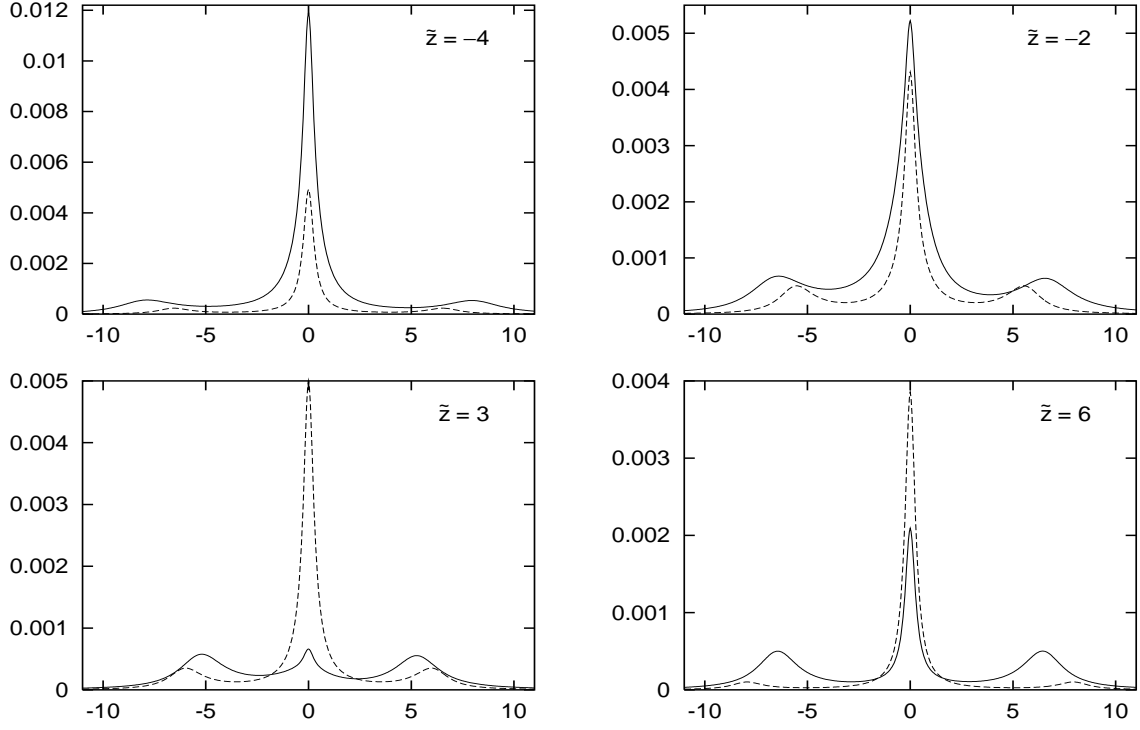}}
\caption{Total spectrum as a function of the frequency $x$ for $\eta^2= 28$,
$\widetilde \gamma=0.6$ and $\widetilde z= -4,\, -2,\, 3,\, 6$; solid line:
$\delta_0^+=-0.03$, $\delta_0^-=0.13$, $\left\| P_\bot g_\pm \right\|^2 =
0.005$, $\left\| P_\bot \Delta g \right\|^2= 0.02$, $\varepsilon/
\|\alpha\|^2=-0.001$; dashed line: $\delta_0^\pm=0$, $\left\| P_\bot g_\pm
\right\|^2 = \left\| P_\bot \Delta g \right\|^2= 0$, $\varepsilon/
\|\alpha\|^2=0$.} \label{fig3}
\end{center}
\end{figure}

Finally, in Fig.\ \ref{fig3} we show some out of resonance spectra (detunings
$\widetilde z = -4,\, -2,\, 3,\, 6$) for $\eta^2 =28$ and the other parameters
as in Figs.\ \ref{fig1} and \ref{fig2} (solid lines); again, the dashed lines
give the Mollow spectrum. Now, a strong difference from the usual case is
shown, consistent with the strong asymmetry in $\widetilde z$ shown by the
total and the elastic cross sections in Fig.\ \ref{fig1}.

%
%

\end{multicols}
\widetext

\end{document}